\documentclass[fleqn]{aa}  

\usepackage{subfig}
\usepackage{graphicx}
\usepackage[font=small,labelfont=bf]{caption}
\usepackage{wrapfig}
\usepackage{subcaption}
\usepackage{booktabs}
\usepackage{float}
\usepackage{txfonts}
\usepackage{pdflscape}
\usepackage{siunitx}
\usepackage[colorlinks=true, citecolor=blue, linkcolor=blue]{hyperref}
\usepackage[utf8]{inputenc}
\usepackage{tabularx}
\usepackage[dvipsnames]{xcolor}
\usepackage{amsmath}
\usepackage{ragged2e}
\usepackage{adjustbox}
\usepackage{amssymb}
\usepackage[normalem]{ulem} 


\newcommand{\angs}{{\rm \AA}}

\newcommand{\sols}{ \rm M_{\odot}}
\newcommand{\teff}{\rm T_{eff}}
\newcommand{\logg}{\rm log\,g_{*}}
\newcommand{\vrad}{\rm v_{rad}}

\usepackage[normalem]{ulem}

\begin{document} 
\date{}

   \title{Carbon measurements in two ultra-faint dwarf galaxies: \\Grus II and Tucana IV}

   \author{Valentina Verdiani \inst{1,2},
           Ása Skúladóttir\inst{1,2},
           Romain Lucchesi \inst{1,2}, 
           Alessio Mucciarelli \inst{3,4},
           Davide Massari \inst{4}, \\
           Giuseppina Battaglia \inst{5,6},
           José María Arroyo-Polonio \inst{5,6},
           Eline Tolstoy \inst{7},
           Sara Covella \inst{3},
           Salvatore Taibi \inst{8}
          }

   \institute{Università degli Studi di Firenze, Via G. Sansone 1, I-50019 Sesto Fiorentino, Italy.
              \\
              \email{valentina.verdiani1@unifi.it}
         \and
             INAF/Osservatorio Astrofisico di Arcetri, Largo E. Fermi 5, I-50125 Firenze, Italy 
         \and 
            Dipartimento di Fisica e Astronomia, Alma Mater Studiorum, Università di Bologna, Via Gobetti 93/2, 40129 Bologna, Italy
         \and 
            INAF, Osservatorio di Astrofisica e Scienza dello Spazio, Via Gobetti 93/3, 40129 Bologna, Italy
         \and
            Instituto de Astrofísica de Canarias, Calle Vía Láctea s/n, 38206 La Laguna, Santa Cruz de Tenerife, Spain
         \and
            Universidad de La Laguna, Avda. Astrofísico Francisco Sánchez, 38205 La Laguna, Santa Cruz de Tenerife, Spain
         \and 
            Kapteyn Astronomical Institute, University of Groningen, PO Box 800, 9700 AV Groningen, The Netherlands
         \and
            Institute of Physics, Laboratory of Astrophysics, École Polytechnique Fédérale de Lausanne (EPFL), 1290 Sauverny, Switzerland
             }

  \abstract
{The ultra-faint dwarf galaxies (UFDs) are some of the oldest and most metal-poor environments in the Local Group. In particular, they are predicted to host the first stars (only H and He) that lit up in our Universe. No metal-free stars have been found to date, but their chemical products can be observed on the surfaces of the ancient second-generation stars such as the carbon-enhanced metal-poor stars (CEMP-no, [C/Fe] $>$ +0.7). However, in each UFD there are only a few stars bright enough for spectroscopic follow-up, therefore it is crucial to study as many of these systems as possible.
Here we follow up stars belonging to two recently discovered UFDs, Grus~II and Tucana~IV.
The spectra analyzed were obtained with the multi-object spectrograph FLAMES/Giraffe at the Very Large Telescope (VLT). This includes spectra in two wavelength ranges: red spectra around the $\ion{Ca}{II}$ triplet (8498\,$\angs$, 8542\,$\angs$, 8662\,$\angs$) used to derive radial velocity and [Fe/H], and blue spectra covering the CH band at $\sim$ 4300\,$\angs$. In total, we analyzed 21 spectra of member candidates for Grus~II and 17 for Tucana~IV, including both Red Giant Branch (RGB) and Horizontal Branch (HB) stars. 
We identified 13 members in Grus~II (thereof 8 RGB stars) and 7 members in Tucana~IV (thereof 3 RGB stars). Among the RGB stars in Grus~II, we found three CEMP-no stars at [Fe/H] $\approx-$3 and [C/Fe]~$>$~+1 and two CEMP-no stars at slightly higher [Fe/H] and [C/Fe]~$>$~+0.7.  In Tucana~IV, we found one CEMP-no star ([Fe/H]~=~$-$2.75 and [C/Fe] = +0.83). 
This project, along with future investigations of CEMP stars in UFDs, allows us to study the impact of the first stars in these ancient and primitive systems and consequently the first chemical enrichment that occurred in the Universe.}

\authorrunning{Verdiani et al.}   
\titlerunning{Carbon measurements}
\maketitle

\section{Introduction}
\label{intro}
The discovery of the ultra-faint dwarf galaxy (UFDs, L $<10^{5}L_{\odot}$, \citealt{Simon2019}) satellites of the Milky Way has piqued a wide interest: these systems are very metal-poor and among the most ancient objects in the Local Group \cite[e.g.][]{salvadori_ultra_2009, Simon2019}. Several studies affirm that the UFDs could be the first mini-halos that hosted the first stars (Pop III, \citealt{bovill_pre-reionization_2009, salvadori_carbon-enhanced_2015}). These predictions are corroborated by the star formation histories of these systems \citep{brown_quenching_2014,gallart_star_2021} that suggest that the star formation only lasted for the first $\sim$Gyr of the Universe.

Pop~III stars are formed from a gas cloud with a primordial chemical composition (H, He). Due to the lack of metals, the roto-vibrational transitions of the H$_2$ molecules are the only cooling channel available (T $\sim 10^4$~K, \citealt{tegmark_how_1997}). Since this cooling channel is not efficient, the first stars are predicted to be more massive than present-day stars \citep[e.g.][]{hirano_one_2014}. In a few Myr, these massive Pop III stars released into the interstellar medium (ISM) the chemical elements heavier than He through supernova (SN) explosions. Thus, Pop~III stars are responsible for the early chemical enrichment of the primordial ISM and play a key role in the onset of the reionization of the Universe \citep[e.g.][]{bromm2011}. To date, the first stars have not been directly observed and we don't know if low-mass long-lived Pop~III stars were able to form since in this case they should be observable in the Local Group \citep[e.g.][]{rossi2021}. 

Even though Pop~III stars have never been found at the present moment, the galactic archaeology studies of UFDs are fundamental to understanding the first chemical enrichment of the Universe since there is a high probability that they can host the descendants of Pop III stars \citep{frebel_chemical_2012}. Due to UFDs low gravitational potential well, the chemical products of the most energetic SNe are unlikely to be retained in the galaxy environment, resulting in a low fraction of metals and consequently a quenched star formation \citep{rossi_hidden_2024}. However, UFDs are able to retain the chemical products of the lowest-energy SNe. The recent discovery of an extremely metal-poor second-generation star in the Pictor II UFD \citep{Chiti2025} supports this hypothesis.

When investigating Pop~III descendants, the carbon-enhanced metal-poor stars (CEMP-no; [C/Fe] > 0.7, with no Ba enhancement) are of special interest: they are the descendants of Pop~III stars exploding as faint supernovae (E $\lesssim$ 10$^{51}$ erg) which pollute the environment mainly with lighter elements, such as C, resulting in very high [C/Fe] ratios \citep{iwamoto_first_2005}. This peculiar chemical pattern will be preserved in the photosphere of the descendant stars which can survive until the present day and are observed both in the Milky Way and its satellites \citep[e.g.][]{aoki_carbonenhanced_2007,  norris_chemical_2010, skuladottir_first_2015}. We note that other formation scenarios have been proposed to explain the CEMP-no stars, such as inhomogeneous internal mixing after the first SNe \citep{2019HY}, the accretion of C-rich gas from a companion in binary systems \citep{2020Komiya}, or metal-pollution due to second-generation star explosions \citep{2021Jeon}.

Another class of CEMP stars is mostly found in binary systems \citep{starkenburg2014,hansenT2015} and they have been enriched both in C and s-process elements (like Ba) by an Asymptotic Giant Branch (AGB) companion. These are called CEMP-s stars \citep[see e.g.][]{Beers2005} and are not representative of the environment in which they were born, i.e. they do not descend from the Pop~III stars. CEMP-s stars are defined as those with [Ba/Fe] $>$ 1 and are typically found at higher [Fe/H] compared to CEMP-no stars \citep{norris_2013}. Additionally, they can be recognized because they present higher absolute carbon abundances, A(C), than CEMP-no stars \citep{bonifacio_topos_2015}. Additionally, we mention the CEMP-r stars class, i.e. the CEMP stars enriched with r-process elements (such as Eu) and CEMP-r/s stars, which are enhanced both in r- and s-process elements. Both of these objects are significantly rarer than CEMP-s and CEMP-no stars.  \citep{Beers2005}.

The fraction of CEMP-no stars increases towards lower metallicities \citep[e.g.][]{Beers2005,lee2013,placco_carbon-enhanced_2014}, and indeed in the metal-poor UFDs the fraction of these Pop III descendants is high \citep{ norris_chemical_2010,salvadori_carbon-enhanced_2015, Ji2020}. The CEMP fraction trend in the Milky Way halo is similar to that of the UFDs. Although several CEMP-no stars have been found in dwarf spheroidal galaxies (dSphs) (e.g. \citealt{skuladottir_first_2015, Susmitha2017}, Cuadra et al. in prep), their fraction seems to be low \citep[e.g.][]{Starkenburg2013A&A...549A..88S}. From low-resolution spectra (R$\approx$2000), \citet{Chiti2018} reported that the CEMP-no fraction in Sculptor is rising towards lower [Fe/H], reaching values similar to the Milky Way. However, this result is not seen in any of the higher resolution studies in the Sculptor dSph \citep{Frebel2010, Tafelmeyer2010, Starkenburg2013A&A...549A..88S, Simon2015, Jablonka2015, skuladottir_first_2015, Skuladottir2021, skuladottir_tracing_2024}. Furthermore, a recent compilation of literature data by \citet{lucchesi_extremely_2024} shows a significantly lower fraction of CEMP-no stars in dSphs. However, the uncertainties are large, suggesting that further investigation is needed, especially at the lowest metallicities.

In UFDs, spectroscopic follow-up is feasible only for the few brightest stars, so statistics are lacking. Therefore, in this work we targeted Red Giant Branch (RGB) and Horizontal Branch (HB) stars in two recently discovered UFDs, Grus~II and Tucana~IV \citep{drlica-wagner_eight_2015}, visible from the Southern Hemisphere.
The stellar masses of the galaxies are M$_*$=3.4$_{-0.4}^{+0.3}\times 10^3\,\rm M_{\odot}$ for Grus~II and M$_*$=2.2$_{-0.3}^{+0.4}\times 10^3\,\rm M_{\odot}$ for Tucana~IV, and their heliocentric distances are respectively d = 53 $\pm$ 5 kpc and d~= 48 $\pm$ 4 kpc \citep{drlica-wagner_eight_2015}. 
Until today, these two UFDs are still poorly studied. \citet{Massari2018} and \citet{PaceLi2019} were the first to determine their kinematic properties and to study their orbit around the Milky Way. \cite{simon_birds_2020} analyzed some stars in these galaxies, deriving the radial velocities (that is their heliocentric line-of-sight velocities), the metallicities and properties of the galaxies, such as their velocity dispersion and orbits around the MW. Additionally, \cite{hansen} determined detailed chemical abundances of the brightest stars in Grus~II. 
However, a derivation of carbon abundances to determine the number of CEMP stars in these two UFDs has never been done. In this work, we derived the line-of-sight velocity, metallicity and carbon abundance measurements of candidate targets of stars belonging to Grus~II and Tucana~IV, aiming to identify member stars and investigate [C/Fe] at low [Fe/H] in these galaxies to identify CEMP stars. 

\begin{figure}[htbp!]
    \centering
        \includegraphics[width=1\linewidth]{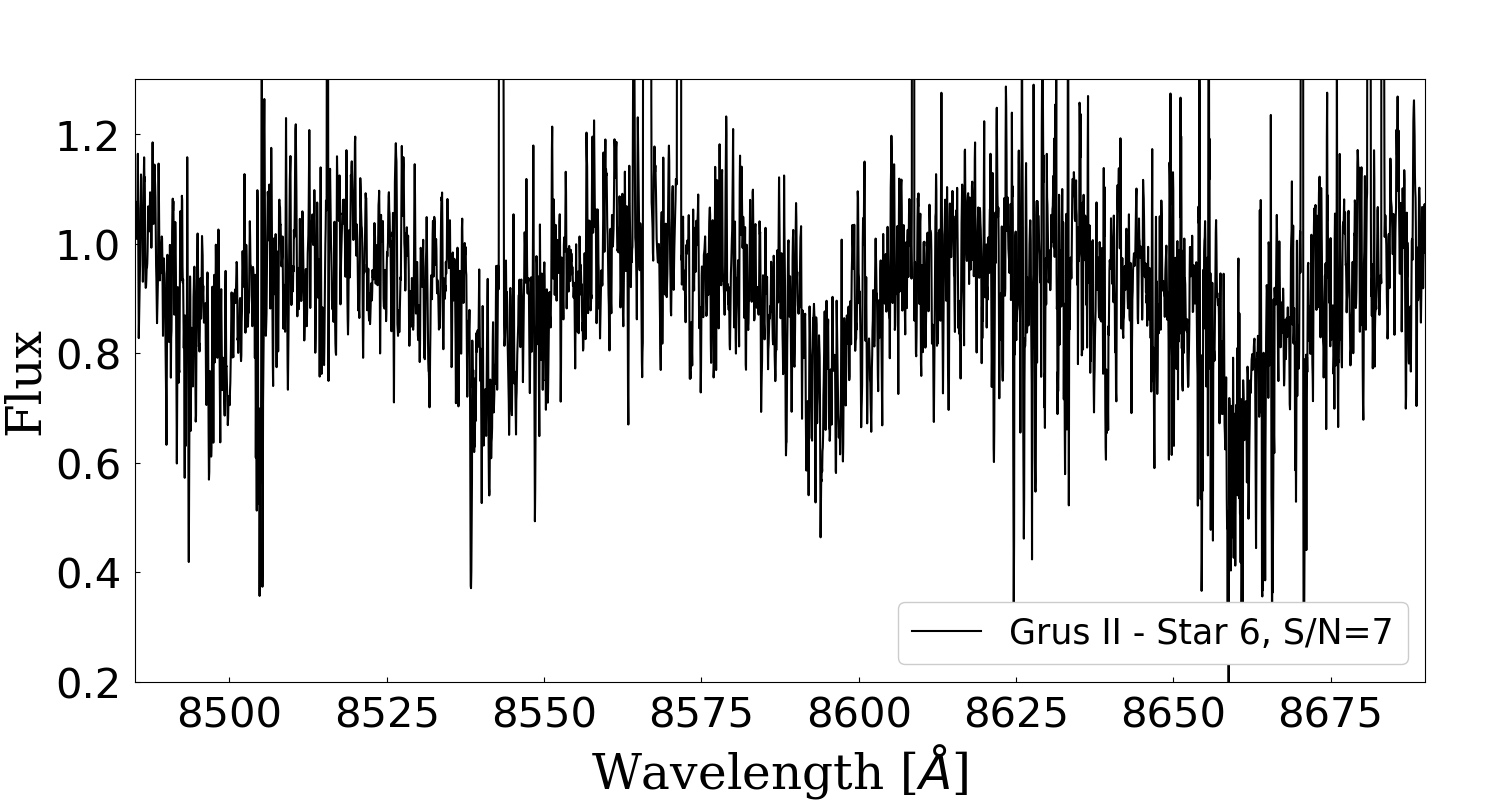}

    \includegraphics[width=1\linewidth]{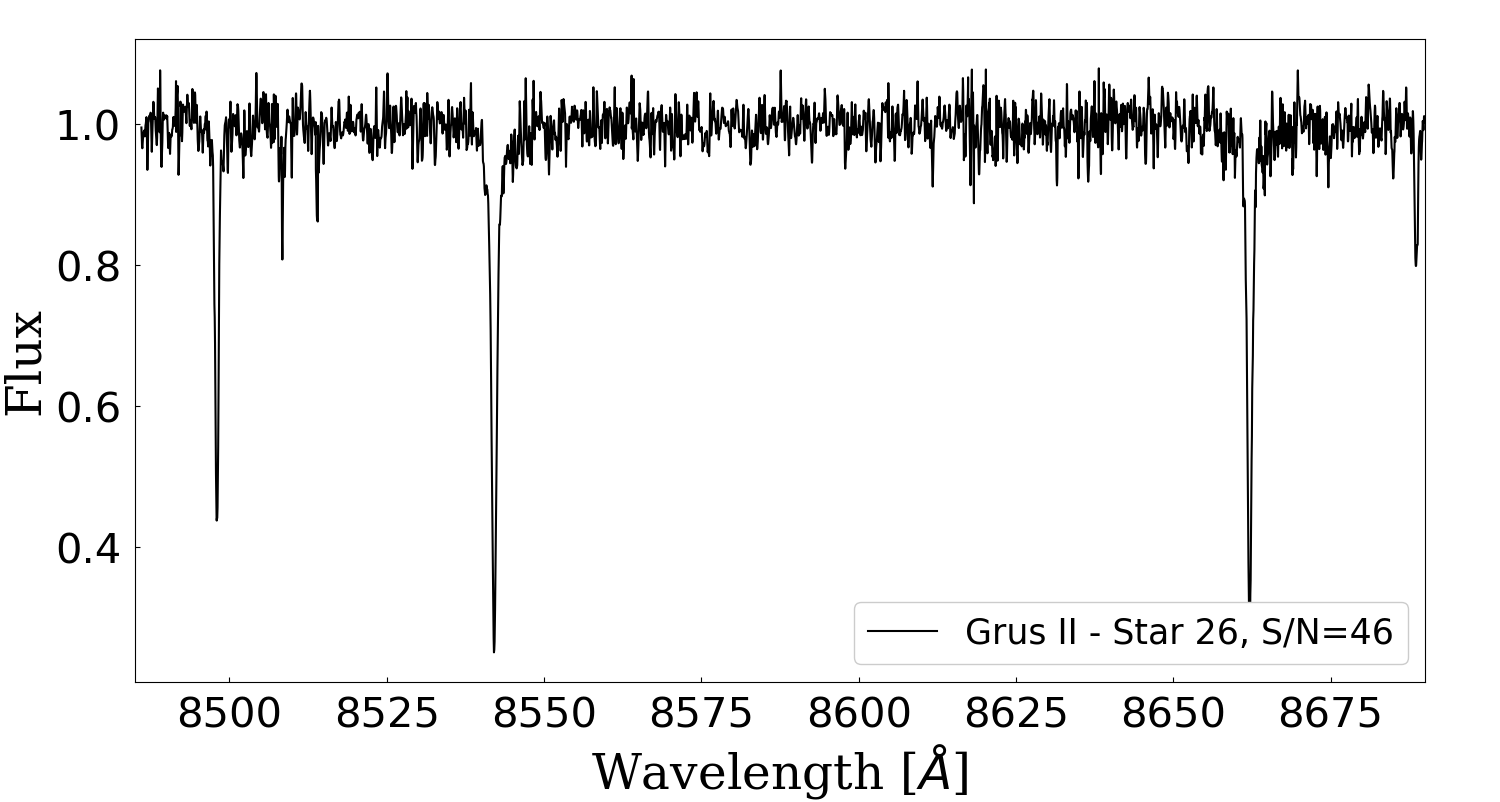}
        \includegraphics[width=1\linewidth]{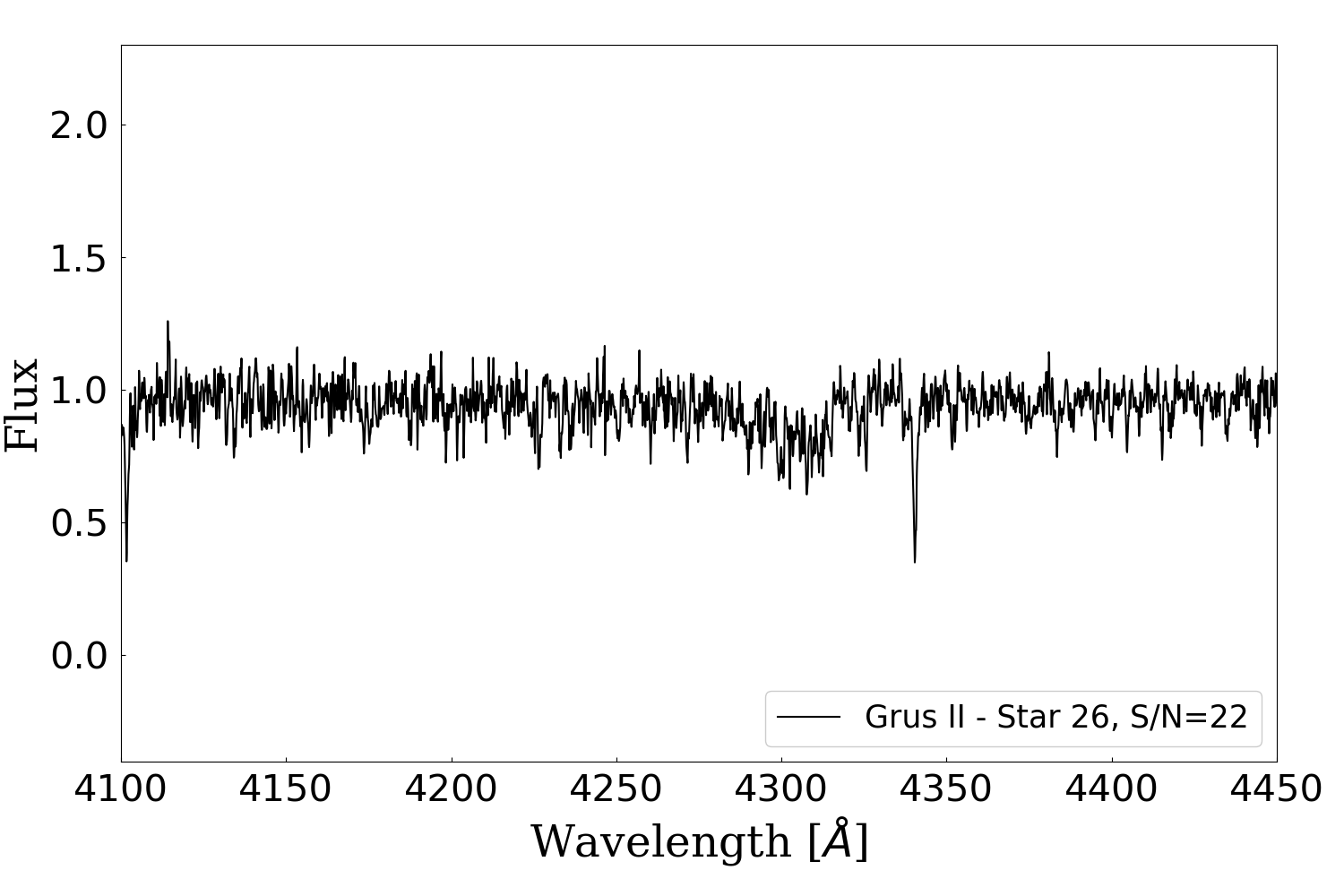}

    \includegraphics[width=1\linewidth]{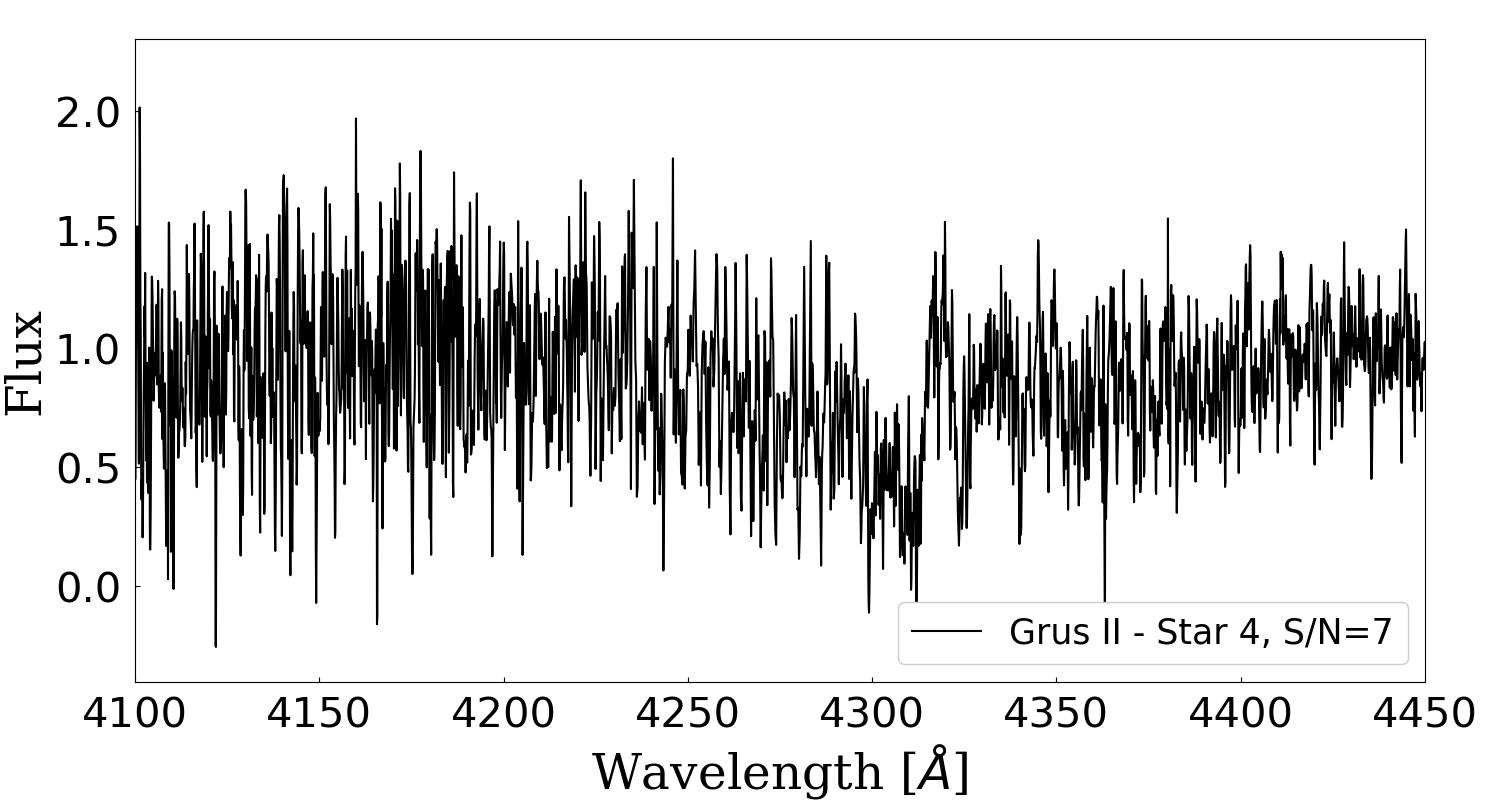}
    \caption{Example of spectra of our target stars. \textbf{Top}: star 6 in Grus~II, a hot HB star whose spectrum is dominated by Paschen lines. \textbf{Middle}: the \ion{Ca}{II} triplet (2nd panel) and the CH band (3rd panel) spectra of the RGB star 26 in Grus~II,   \textbf{Bottom}: CH band spectrum of the RGB star 4 in Grus II.}
    \label{RGB and HB spectra}
\end{figure}

\begin{figure*}[t]
    \hspace{-1em}
    \includegraphics[width=0.55\linewidth]{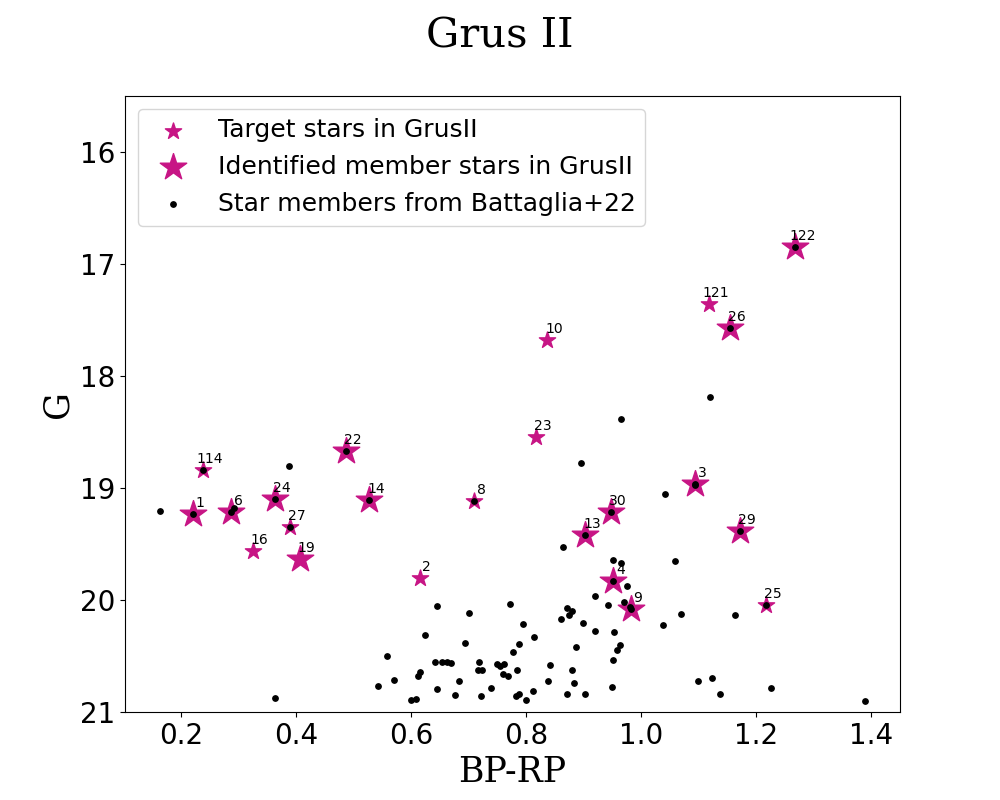}
    \hspace{-3em}
    \includegraphics[width=0.55\linewidth]{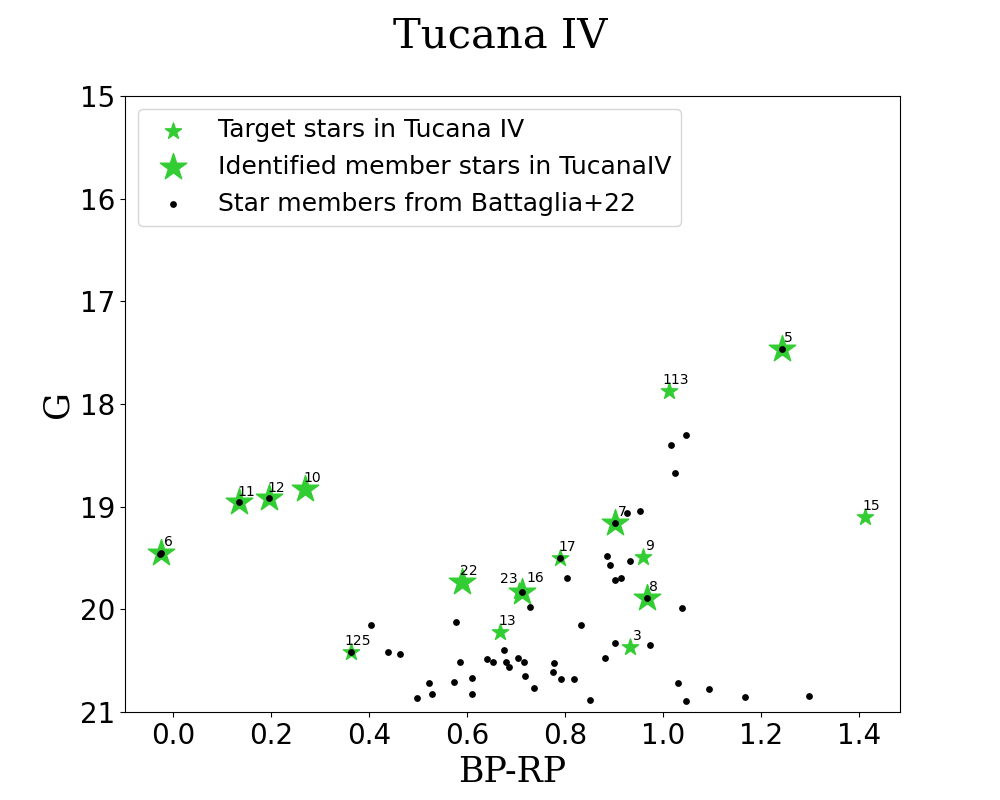}
  \caption{Color-magnitude diagram of Grus II (left panel) and Tucana IV (right panel). The star symbols indicate our FLAMES/GIRAFFE targets, larger symbols are the member stars determined in Section \ref{vrad text}, and black small circles represent possible members from the catalog by \cite{battaglia_gaia_2022} based on Gaia eDR3 measurements. 
  Numbers are the star IDs (\autoref{tab par G} and \autoref{tab par T}). \autoref{isochrones} also shows the CMD with isochrones of different metallicities overplotted.}
  \label{color mag}
\end{figure*}

\section{Observational data}
\label{obs}
We selected the spectroscopic targets based on the mean proper motion of the two galaxies as found by \cite{PaceLi2019}. Candidate members were chosen among bright stars having a proper motion consistent with the mean galaxy motion within 3$\sigma$.
The target stars were observed with the multi-object spectrograph FLAMES/Giraffe at the Very Large Telescope (VLT; ESO program IDs 0103.B-0163, 105.205T, PI D. Massari). The HR21 setting was used to cover the \ion{Ca}{II} triplet region (8498~$\angs$, 8542~$\angs$, 8662~$\angs$) in the red, while the CH molecular band ($\sim$4300~$\angs$) in the blue was covered by the LR2 grating (see \autoref{tab obs}). The exposure time of each observing block is 3600~s.
Every star in Tucana IV got 4h of observations for both the red and blue, while Grus II potential members got either 2h or 4h (see \autoref{tab SN gru}). The first observation was carried out in July 2019, the second from July to September 2021. 
\autoref{RGB and HB spectra} shows examples of red and blue spectra in Grus~II.
The sample contains both stars belonging to the RGB and HB (\autoref{RGB and HB spectra}). For each exposure obtained with HR21, we cross-correlated the position of the observed sky emission lines against a sky spectrum selected among the sky spectra of our dataset. This check was performed for each individual exposure before co-adding together the spectra of the same target. We found typical differences (in absolute value) smaller than 0.4~km~s$^{-1}$, corresponding to one-fourth of the GIRAFFE pixel. Moreover, no significant systematic offset was found, as the average difference was +0.05~km~s$^{-1}$ with a $\sigma$ of 0.14~km~s$^{-1}$. We also checked possible fiber-dependent effects like those mentioned in \cite{Jenkins2021}, but we did not find any significant patterns.

The proper motions (pmra and pmdec, in this work $\rm \mu_{\alpha}cos(\delta)$ and $\mu_{\delta})$, the right ascension (RA), the declination (DEC) in J2000, the magnitudes (phot$\textunderscore$g$\textunderscore$mean$\textunderscore$mag, phot$\textunderscore$bp$\textunderscore$mean$\textunderscore$mag, phot$\textunderscore$rp$\textunderscore$mean$\textunderscore$mag, in this work G, BP, RP) and other parameters were taken from the Gaia Data Release~3 (Gaia DR3) and are listed in \autoref{tab par G} and \autoref{tab par T}. \autoref{color mag} shows the color-magnitude diagram (CMD) of the spectroscopic sample, along with likely members based on Gaia eDR3 from the selection of \cite{battaglia_gaia_2022}. Moreover, in Appendix \ref{Isochrones}, we show the CMD with the isochrones overplotted (\autoref{isochrones}).

The spectra were reduced using the ESO pipeline developed for FLAMES-GIRAFFE spectra\footnote{https://www.eso.org/sci/software/pipe$\textunderscore$aem$\textunderscore$main.html}. The pipeline workflow performs for each individual observation bias-subtraction, flat-fielding, wavelength calibration using a standard Th-Ar lamp, and spectral extraction. The accuracy of the wavelength calibration for the spectra observed with HR21 was checked by measuring the position of several sky lines (not available in the LR2 spectra). For each exposure, the stellar spectra were corrected for the sky background contribution by subtracting a spectrum obtained by median-averaging together all the individual sky spectra. After performing the heliocentric corrections, the spectra were normalized and co-added with the removal of cosmic rays.

After this procedure, we determined the signal-to-noise ratio (S/N) by deriving the variance of the flux in line-free regions. In the case of the blue spectra, the S/N was difficult to measure due to lack of line-free regions (see \autoref{RGB and HB spectra}, bottom panel), and the listed S/N in the blue should therefore be considered as a lower limit (\autoref{tab SN gru}).

\begin{table}[t]
\caption{Observational settings.}
\label{tab obs}
\renewcommand{\arraystretch}{1.4}
\centering
{\large
 \begin{tabular}{ccc}
 \hline
 FLAMES setting & Resolution & $\lambda$ [$\angs$]\\
 \hline\hline
   LR2 & 6\,000 & 3960 - 4570\\
   HR21 & 18\,000 & 8480 - 8980 \\
 \hline
 \end{tabular}
}
\end{table}

\section{Atmospheric parameters and stellar models}
\label{param}
\subsection{Effective temperature}
To derive the effective temperatures the following formula was used \citep{mucciarelli_gaia_2020}: 

\begin{equation}
    \rm T_{\rm eff}=\dfrac{5040}{\theta}
\end{equation}
with $\theta$ defined as:
  \begin{equation}
  \label{tmuc}
\theta=b_0+b_1C+b_2C^2+b_3\text{[Fe/H]}+b_4\text{[Fe/H]}^2+b_5\text{[Fe/H]}\,C
  \end{equation}
where $b_0...b_5$ are the parameters for giant stars from \cite{mucciarelli_exploiting_2021} and $C$ represents the colors (BP$-$RP)$_0$, (BP$-$G)$_0$, (G$-$RP)$_0$, obtained from the magnitudes in the G, BP and RP bands provided by Gaia. This calibration is not adequate for hot stars ($\teff$ $>$ 6500~K) since \autoref{tmuc} holds for RGB stars, but it was used in this work with the only aim of identifying the HB stars. The spectra were then confirmed as belonging to hot stars by checking for the presence of strong Paschen lines (see \autoref{RGB and HB spectra}, top panel). Moreover, we compared the position of the stars in the color-magnitude diagram (\autoref{color mag} and \autoref{isochrones}) with our results to confirm a reliable identification of HB stars, all having (BP$-$RP)~$\lesssim$~0.5. 
The colors are extinction and reddening corrected: the extinction factor is
\begin{equation}
    A_{X}=A_{0}k_{X}
\end{equation}
where $X$ are the magnitudes (G, BP, RP) and $k_{X}$ is defined by Equation~1 in \cite{gaia_collaboration_gaia_2018}. 
The color-independent extinction coefficient is provided by Gaia: $A_0~=~3.1E(B-V)$, where $E(B-V)$ is the color excess. The term $(\text{BP$-$RP})_{0}$ that appears in the definition of $k_{X}$ was derived by performing four iterations, with $(\text{BP$-$RP})_{0}$ equal to $(\text{BP$-$RP})$ measured directly by Gaia as the initial guess. The results were used to derive the $A_{X}$ coefficients and therefore the real magnitudes and colors. 

The final adopted $ \rm T_{\rm eff}$ is the average of the three $ \rm T_{\rm eff}$ obtained from the three colors. In the Appendix, \autoref{tab par G} and \autoref{tab par T} report the corrected colors and the mean effective temperature of every star of the sample.
The error on the $\rm T_{eff}$:
\begin{equation}
    \sigma_{\rm T_{eff}}=\sqrt{\sigma_1^2+\sigma_2^2}
\end{equation}
where $\rm \sigma_1=\dfrac{\sigma_{std}}{\sqrt{2}}$ is the error on the mean of the three color temperatures, while
 $\rm \sigma_2=\dfrac{s_1+s_2+s_3}{3}$ is the mean of the three errors associated to the colors ($\rm s_1=83$~K, $ \rm s_2=83$~K and $\rm s_3=71$~K.), which represents the intrinsic systematic error of the estimated method from \cite{mucciarelli_exploiting_2021}.

\subsection{Surface gravity}
The surface gravity is evaluated using the known Stefan-Boltzmann relation \citep[see e.g.][]{skuladottir_tracing_2024}:

\begin{equation}
    \rm {log \,\, g}_* = \rm \log g_{\odot} + \rm{log}\dfrac{\rm{M}_*}{\rm{M}_{\odot}} + 4\, \log \dfrac{\rm{T}_{\rm eff}}{\rm{T}_{\rm eff,\odot}} + 0.4\,(\rm{M_{\rm{bol},*}}- \rm{M_{\rm{bol},\odot}})
\end{equation}
where  $\rm M_{\rm{bol},*}=G_0 - (m - M) + BC$ is the bolometric magnitude and BC is the bolometric correction (derived using Eq. 7 in \citealt{2018adrae}). The stellar mass considered is the typical one of the RGB stars in UFDs, i.e. 0.8 $\pm$ 0.2 $\sols$. The solar values used are $\rm \log g_{\odot}$ = 4.44, $\rm{T}_{\rm eff,\odot}$~ = 5772 K and $\rm{M_{\rm{bol},\odot}}$ = 4.74.

The error on the $\logg$ is found assuming negligible errors on the solar parameters: 
\begin{equation}
   \rm \sigma_{\logg}=\sqrt{\sigma^2 \left(\rm log\dfrac{\rm M_*}{\sols}\right) +  4\,\sigma^2\left(\rm log\dfrac{\rm T_{eff,*}}{\rm T_{eff,\odot}}\right) + 0.4\, \sigma^2(\rm M_{bol,*})}
\end{equation}

\subsection{Microturbulence velocity}
From \citet{anthony-twarog_lithium-rich_2013}, we used the empirical relation:
\begin{equation}
   \rm  v_{turb}=2.0 -0.2 \, \logg
\end{equation}
 The error on the v$_{turb}$ is found with 
\begin{equation}
    \rm \sigma_{\rm v_{turb}}=\sqrt{(0.2\,\sigma_{\logg})^2+ \sigma_{\rm v_{turb,rms}}^2}
\end{equation}
where $ \sigma_{\rm v_{turb,rms}}=0.17$~ km~s$^{-1}$, which represents the average scatter between the empirical relation and the measurements \citep{carretta_iron_2004}.

\subsection{Stellar models and synthetic spectra}
\label{model}
Synthetic spectra are generated by using a grid of Model Atmospheres with a Radiative and Convective Scheme (MARCS) \citep{gustafsson_grid_2008} combined with Turbospectrum \citep{2012ascl.soft05004P}. The grid of MARCS 1D atmosphere models is downloaded with the {\it Standard composition}, that is, including the classical $\alpha$-enhancement of +0.4~dex at low metallicity, from the MARCS web site\footnote{\url{http://marcs.astro.uu.se}} \citep{gustafsson_grid_2008}, and interpolated using Thomas Masseron's $interpol\_modeles$ code for the given parameters of each star.
Synthetic spectra are computed for each model, starting from solar chemical abundance ratios ([X/Fe]=0) and an estimated metallicity of $-$2.0. This first estimate of [Fe/H] is used to generate the synthetic spectrum for the derivation of the radial velocity, which is not sensitive to the exact assumed [Fe/H].

Employing the $\chi^2$ minimization method, we find the best fit to an observed spectrum to derive the radial velocities (Section \ref{vrad text}) and chemical abundances (Section \ref{chem text}).

\section{Membership determination}
\subsection{Radial velocities}
\label{vrad text}

\begin{figure}[H]
\hspace{-2em}
\includegraphics[width=1.05\linewidth]{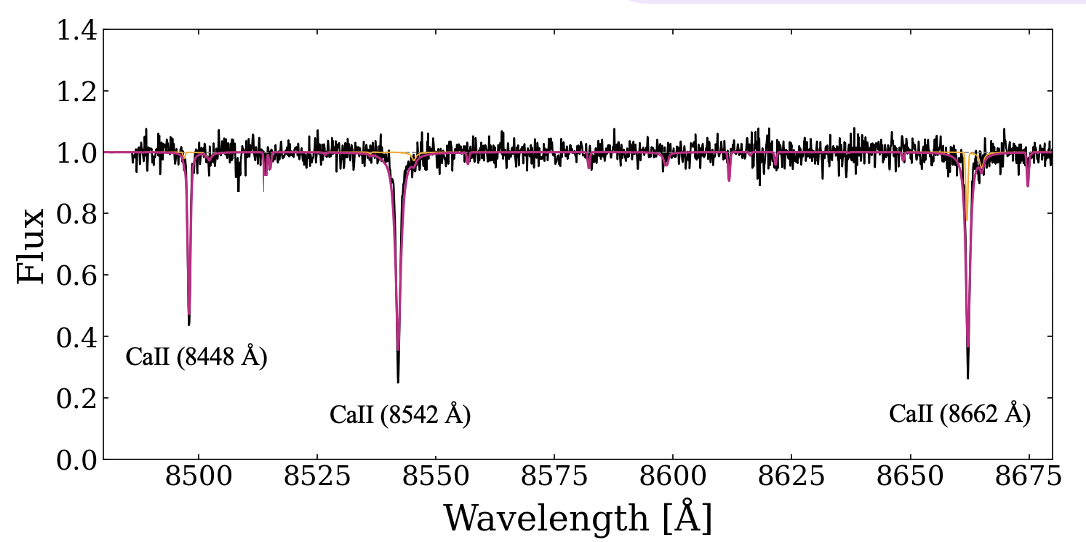} 
\caption{The \ion{Ca}{II} triplet of the star 26 in Grus~II. The black line is the observed spectrum, the magenta line is the synthetic spectrum best fit to the radial velocity of the star.}
\label{Ca synth}
\end{figure}

\begin{figure*}[t]
    \hspace{-1em}
    \includegraphics[width=0.55\linewidth]{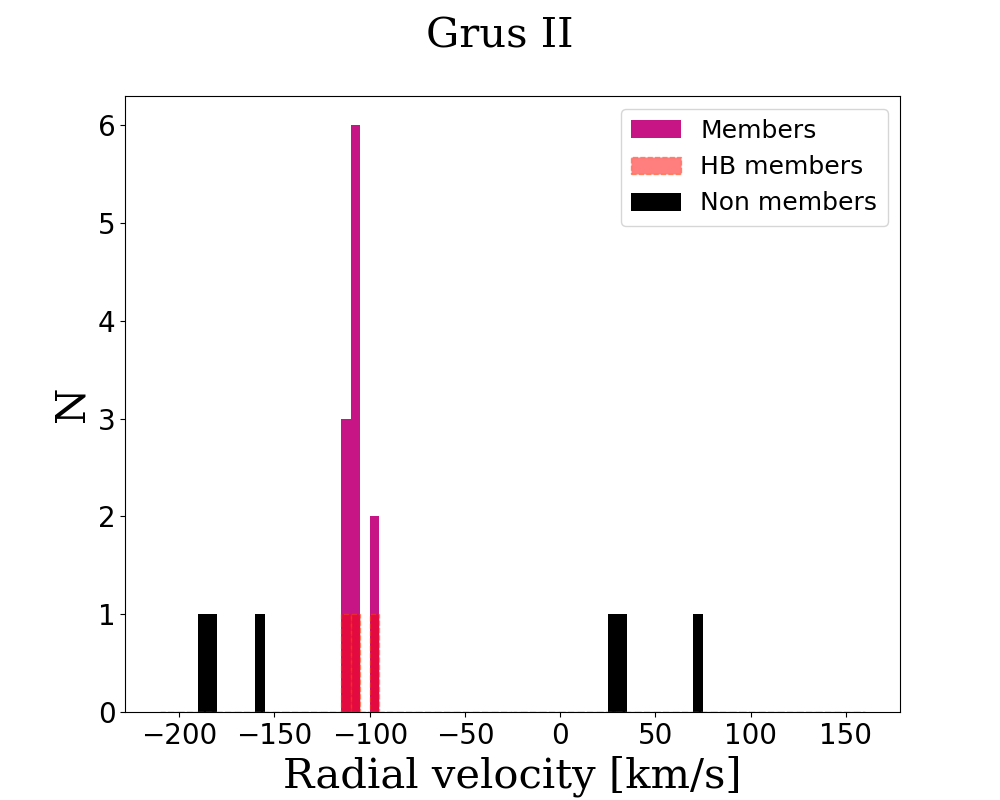}
    \hspace{-3em}
    \includegraphics[width=0.55\linewidth]{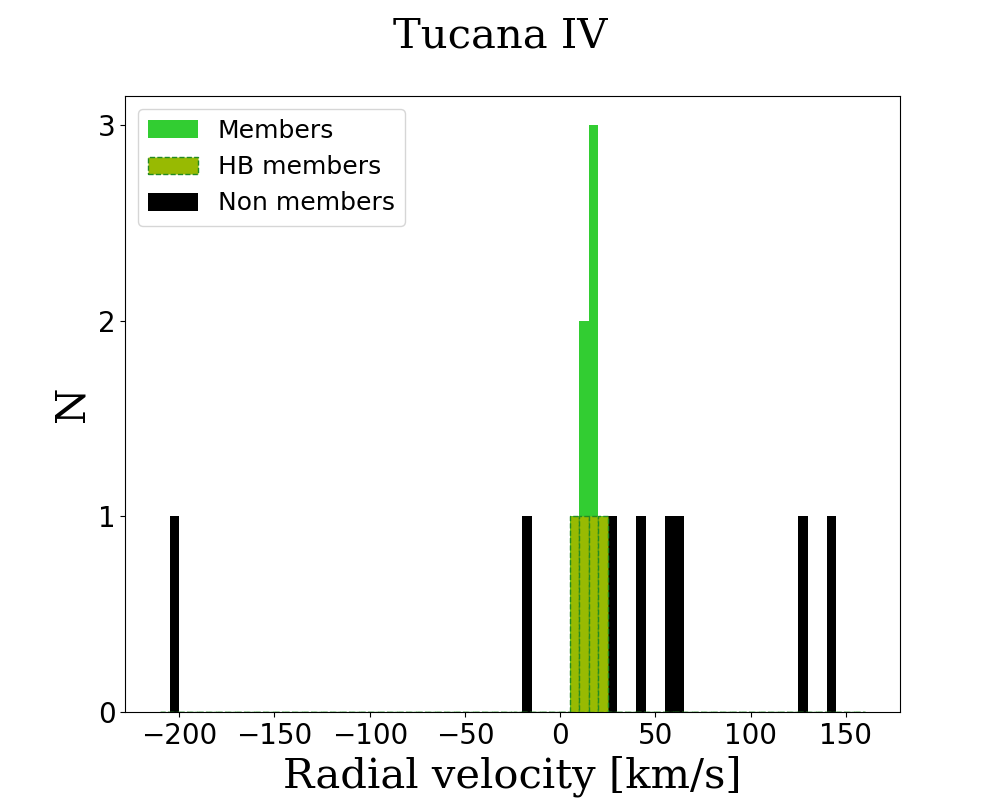}
  \caption{Radial velocity distribution of Grus II (left panel) and Tucana IV (right panel). The black bars represent the non-members, the colored ones are the members. The shaded bars are the HB members, which do not present any significant difference or offset from the RGB stars in radial velocity. We only include stars with red spectra available, but we note that the HB stars 1 and 14 in Grus~II are members based only on measurements from blue spectra (\autoref{tab SN gru}).}
  \label{vrad}
\end{figure*}

After an accurate continuum level identification that allows us to normalize the stellar spectra, our aim was to determine the members among the target stars. We derived the radial velocity ($\vrad$) from the $\ion{Ca}{II}$ triplet lines using the minimization of the $\chi^2$ method, i.e. iterating by shifting the observed spectrum with different values of $\vrad$ in a range between $\pm$100 km~s$^{-1}$ around the average of the galaxies \citep{mcconnachie_observed_2012} in steps of 0.1~km~s$^{-1}$ until the best fit to a synthetic spectrum (created adopting the stellar parameters calculated in Section \ref{param}) is obtained (\autoref{Ca synth}). \autoref{vrad} shows radial velocity distributions of Grus~II and Tucana~IV derived from the red spectra.

\begin{figure*}
    \centering
   \hspace*{-3em} 
   \begin{minipage}[t]{0.5\linewidth}
   \includegraphics[width=1.15\linewidth]{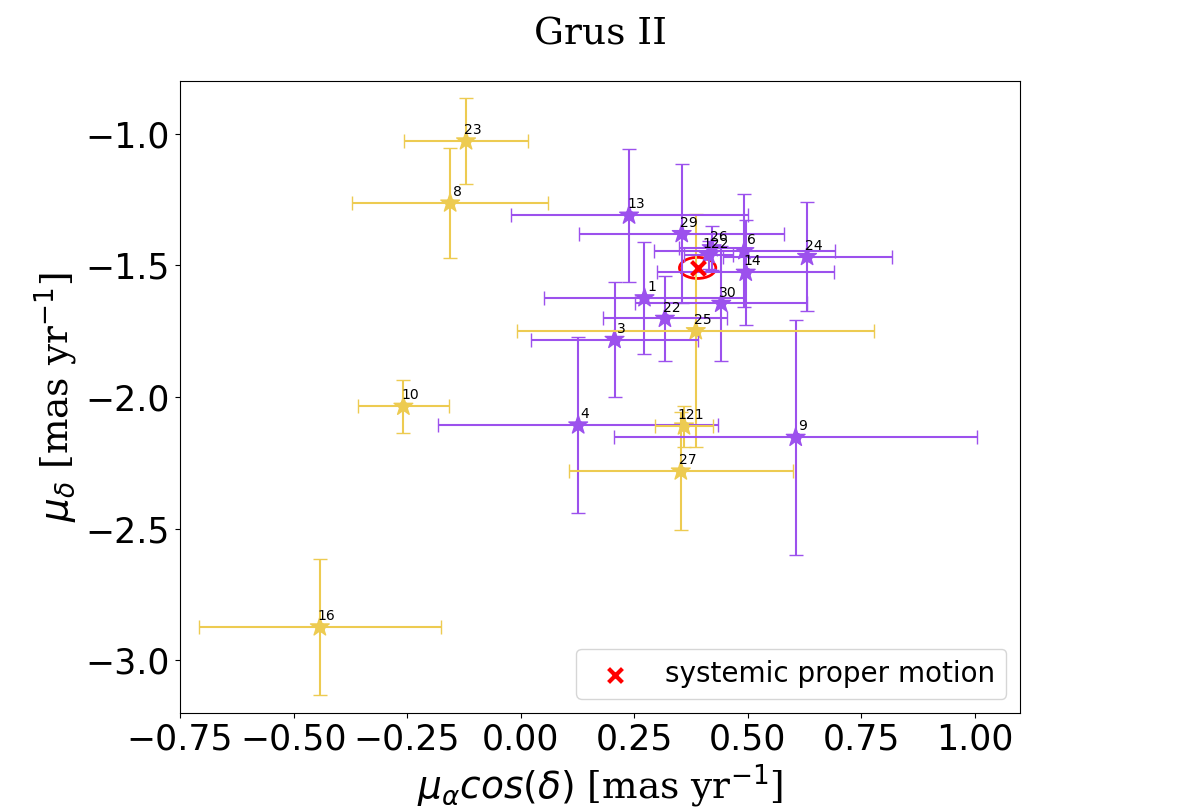} 
   \end{minipage}
   \hspace*{-1em} 
   \begin{minipage}[t]{0.5\linewidth} \includegraphics[width=1.15\linewidth]{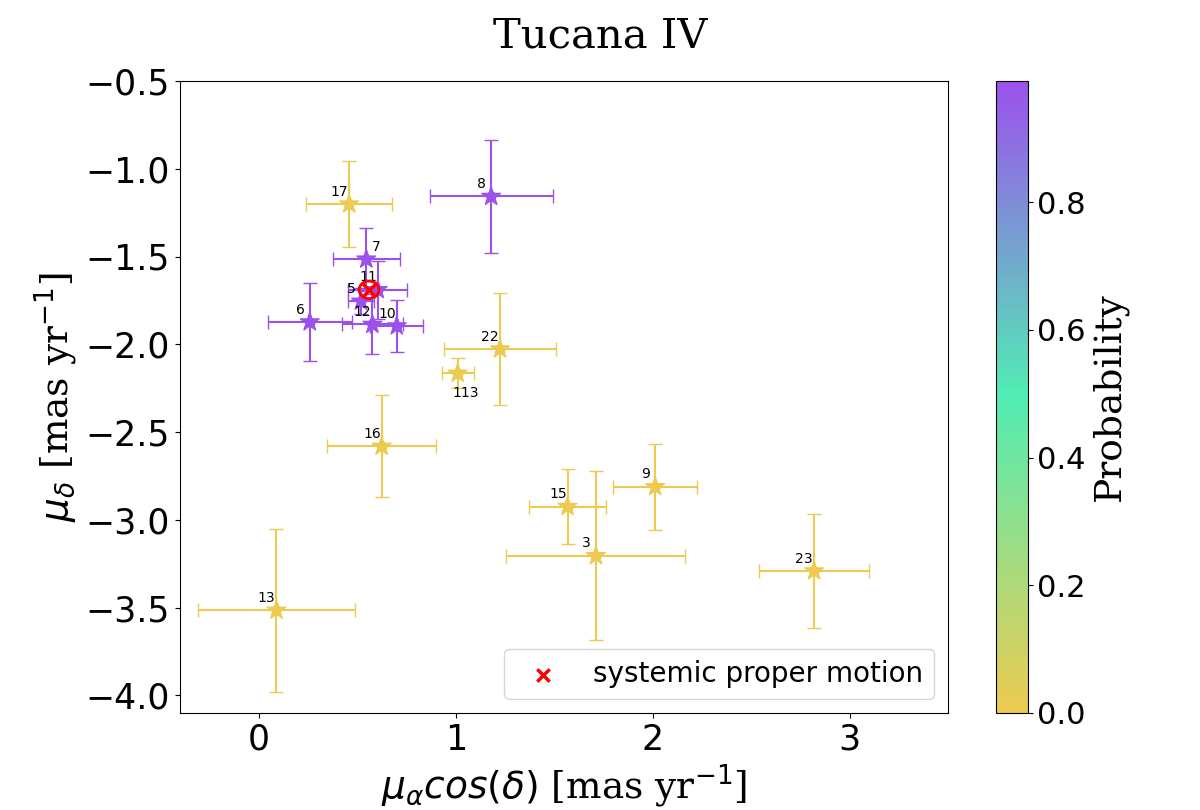} 
   \end{minipage}
    \caption{Proper motions of the stars in Grus~II and Tucana~IV from Gaia DR3. The stars are color-coded according to the membership probability based on \cite{battaglia_gaia_2022}. The red cross is the systemic proper motion \citep{battaglia_gaia_2022}, the red ellipse corresponds to 1$\sigma$, which is the error on the mean proper motions.}
    \label{proper motion}
\end{figure*}

\begin{figure*}
    \hspace{-2em}
    \includegraphics[width=0.57\linewidth]{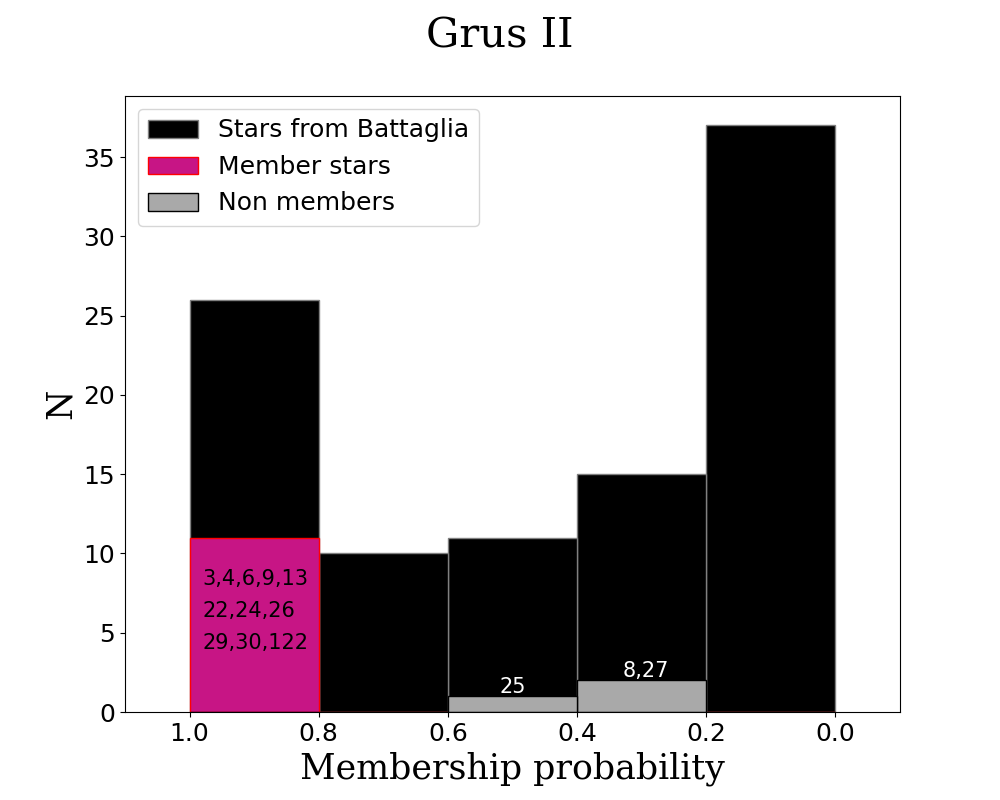}
    \hspace{-3em}
    \includegraphics[width=0.57\linewidth]{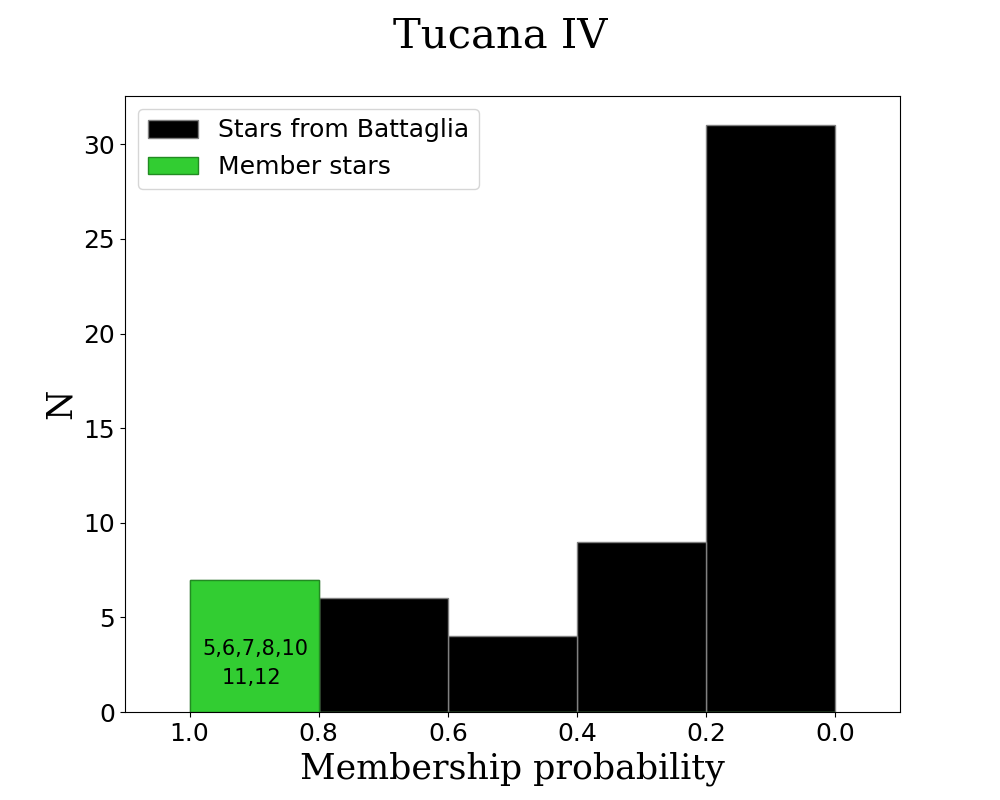}
    \caption{Histograms of membership probability in Grus~II and Tucana~IV from \cite{battaglia_gaia_2022}, showing our targets in colors (members) and gray (our non-members). We note that non-member stars 10, 16, 23, 121 in Grus~II and 3, 13, 15, 23 in Tucana~IV are not included in the \cite{battaglia_gaia_2022} catalog.} 
    \label{memberhist}
\end{figure*} 

\begin{figure*}
\centering
\hspace*{-1.7em}
\begin{minipage}[t]{0.5\linewidth}
\includegraphics[width=1.05\linewidth]{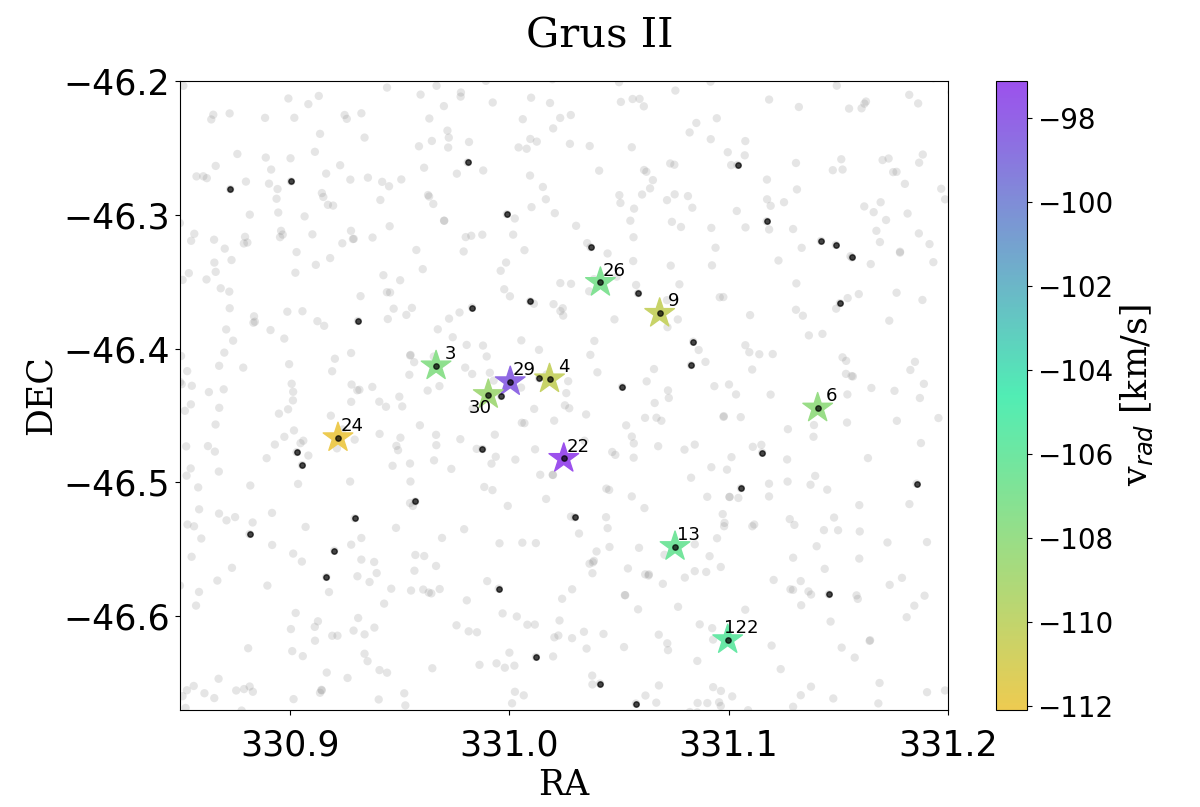} 
 \end{minipage}   
\begin{minipage}[t]{0.5\linewidth}
\hspace{1.5em}
\includegraphics[width=1.05\linewidth]{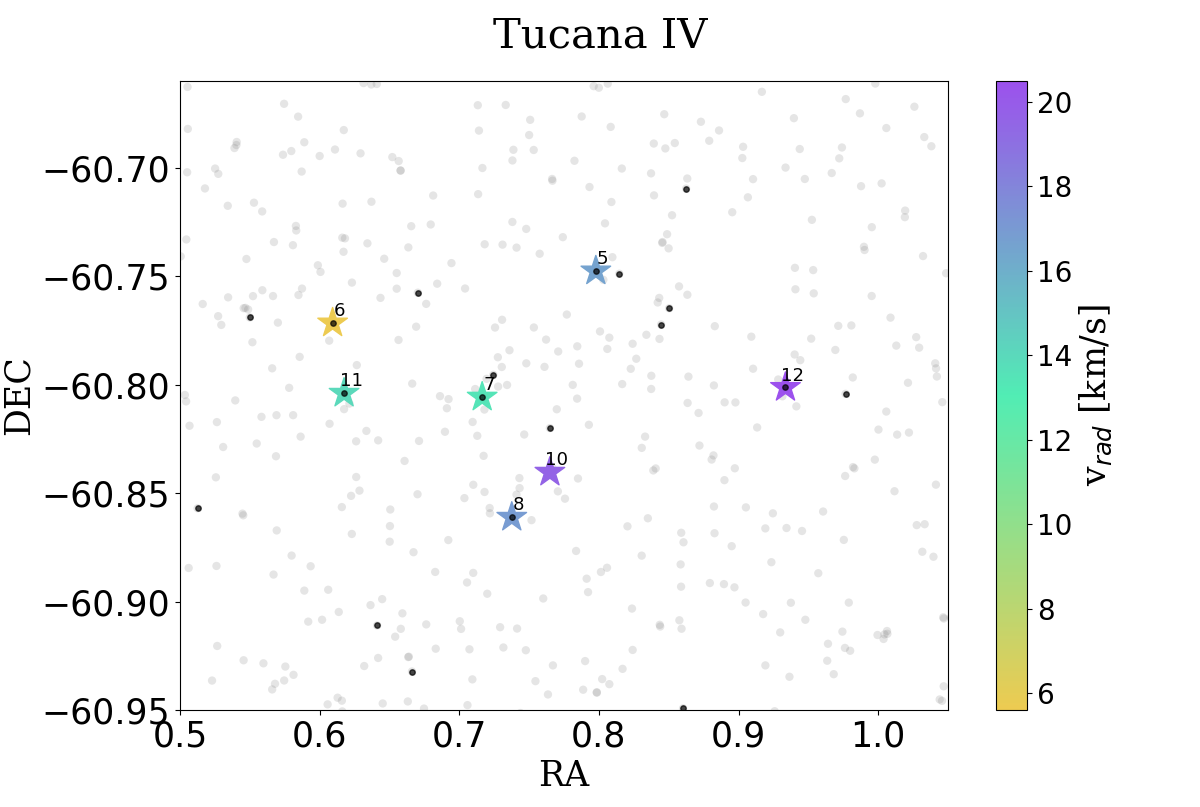}
 \end{minipage}
\caption{Positions of our member stars color-coded by radial velocity.}
\label{ra dec vrad}
\end{figure*}

The radial velocities of every targeted star we analyzed in Grus~II and Tucana~IV are listed in Table \ref{tab SN gru} with corresponding uncertainties. Two stars (2 and 19) in Grus~II were discarded from our analysis since their S/N was very low, preventing us from obtaining reliable measurements. Stars 1, 14 and 114 in Grus~II and star 125 in Tucana~IV do not have any available exposures in the red, therefore their radial velocities are only derived from the blue LR2 setting.

The errors on the radial velocity are found by performing Montecarlo simulations. Using a template spectrum, we made 1000 iterations by adding random Gaussian noise to obtain mock observed spectra at five fixed S/N (5, 15, 30, 50, 70 pix$^{-1}$). After defining an array of radial velocity in steps of 0.1~km~s$^{-1}$, the best value of $\vrad$ is found for all the Montecarlo spectra, comparing them with the best synthetic spectra of the stars in the sample. The distribution of the best values should be centered around 0, but the dispersion increases with lower S/N. Additionally, we tested this method using spectra with different stellar parameters and we obtained consistent results.
The sigma of the distributions derived for these 5 fixed S/N values serves as a reference for performing a polynomial fit to the curve of the radial velocity error as a function of the S/N (typically $\sigma_{\rm v_{rad}}$~$\lesssim$~0.4~km~s$^{-1}$ for S/N > 20~pix$^{-1}$, $\sigma_{\rm v_{rad}}$ $\sim$ 1.5 km~s$^{-1}$ at S/N~$\sim$~10 pix$^{-1}$ and $\sigma_{\rm v_{rad}}$~$\sim$~2.5~km~s$^{-1}$ at S/N $\sim$ 5 pix$^{-1}$). Moreover, star 22 in Grus~II is a special case since the Ca triplet lines are completely contaminated by the skylines: we derived the radial velocities of the two exposures using the Paschen line at $\sim$8600$\AA$. Since this is a broad line, we expected large uncertainty and indeed obtained an error of 5 km~s$^{-1}$ by visual inspection of the comparison between the observed and template spectra.
Two separate polynomial fits were done for the RGB and HB member stars, since the spectra have different features (\autoref{RGB and HB spectra}).

Additionally, we derived the $\vrad$ from the CH-band region for all the stars with LR2 spectra, obtaining the same member identification given by the red spectra. However, these measurements are less reliable since the spectra are of lower resolution (see Table \ref{tab obs}), the S/N is typically lower than the red spectra and the CH-band contains many broad molecular features. Moreover, the radial velocity from broad lines, H$_{\delta}$ and the H$_{\gamma}$ are less trustworthy than the ones obtained with the $\ion{Ca}{II}$ triplet. Furthermore, while the wavelength calibration in HR21 was checked by the positions of the location of the sky lines, there are no sky lines available in the blue LR2 region. This is why the preferred procedure involves the narrower atomic lines of the $\ion{Ca}{II}$ triplet. The same procedure was used for the errors in the blue as for the red. However, since there are no skylines, the zero-point could not be verified, and therefore the systematic errors are not included in the estimated uncertainties. \autoref{diff red blue} compares the results of the radial velocity derived from the red and blue spectra.

\subsection{Combining radial velocities with Gaia}
\label{prob}
For the final member identification, we combined our measured radial velocities with information from Gaia.
\autoref{proper motion} shows the proper motion in RA and DEC of the stars color-coded by the probability of membership and the average systemic proper motion of the galaxy from \cite{battaglia_gaia_2022}. 
In \autoref{memberhist} are reported the histograms of the membership probability of our stars according to \citep{battaglia_gaia_2022}. The colored bars are the identified member stars, while the gray bars are the non-members.
\autoref{ra dec vrad} shows the positions of the stars color-coded by the radial velocity derived in Section \ref{vrad text}. All these results are discussed in detail in the following subsections. 

\subsubsection{\textit{Grus~II}}
Since the radial velocity distribution of Grus~II shows a large velocity offset with respect to Milky Way stars (see \autoref{vrad}), the eleven members we identified in Grus~II are clearly separated from the non-members. The systemic radial velocity of the galaxy is obtained by calculating the weighted mean of the radial velocity of the individual member stars, $\rm v_{sys}~=~-$~106.3~$\pm$~0.1~km~s$^{-1}$, while the velocity dispersion is $\sigma$~=~4.7 $\pm$ 1.0~km~s$^{-1}$.
After the derivation of radial velocity and establishing the star members belonging to the galaxies, we compared our results with the membership probability based on Gaia eDR3 only, reported in \cite{battaglia_gaia_2022}, see \autoref{memberhist}. According to this catalog, all of Grus~II star members we found have a probability higher than 80$\%$ of belonging to the galaxy. The non-member stars, on the other hand, exhibit a probability lower than 60$\%$, (\autoref{memberhist}). We note that the HB star 22 shows some signature of variability in radial velocity (see Section \ref{app vrad blue}).

We note that stars 1, 14 and 114 (all HB stars) have only blue spectra available (see Section \ref{vrad text}). Therefore we derive the radial velocities from the CH-band: the values are $\vrad~=~-$98.8~$\pm$~0.5~km~s$^{-1}$, $\vrad=-$104.0~$\pm$~1.2~km~s$^{-1}$ and $\vrad=-$22.4~$\pm$~1.9~km~s$^{-1}$ respectively. The proper motions of stars 1 and 14 ($\rm \mu_{\alpha}cos(\delta)=~0.273~\pm~0.221$ and $\mu_{\delta}=-1.624~\pm~0.213$  and $\rm \mu_{\alpha}cos(\delta)=~0.496~\pm~0.195$ and $\mu_{\delta}=-1.527~\pm~0.198$, respectively) and their position on the CMD (\autoref{color mag}) support the identification as members, while star 114 is classified as a non-member since its radial velocity differs $>$5$\sigma$ from the peak of the distribution. Therefore, we have confirmed in total 13 stars as members in Grus~II. However, to ensure self-consistency of our results, we do not include these stars (with only blue spectra) in \autoref{vrad} and \autoref{proper motion} or in the following analysis. 
The proper motions of the member stars of Grus~II are shown in \autoref{proper motion} and they are all consistent with the systemic proper motions. Moreover, we do not see any velocity gradient that indicates a tidal disruption of the galaxy (see \autoref{ra dec vrad}).

\subsubsection{\textit{Tucana~IV}}

The mean radial velocity of Tucana~IV is relatively close to zero (see \autoref{vrad}), leading to similar radial velocity between members and non-members due to Galactic contamination. Moreover, we do not know the impact of binary stars, therefore we did not exclude a priori the stars that differ less than 3$\sigma$ from the center of the distribution (stars 9, 16, 17, 22 and 113), where here $\sigma$ is the standard deviation of the distribution considering the certain and potential members identified in this part of the analysis.
We found 12 stars within 50~km~s$^{-1}$ from the average radial velocity of Tucana IV. Here, we are performing a very broad membership assignment based only on the radial velocity information with the aim of not excluding potential members in binary systems. This will now be refined to include also the probabilities of membership determined from Gaia, since we are aware that not all of the considered stars are binaries.
Star 125 has only the blue LR2 exposures available, from which we obtain $\vrad=-$40.0~$\pm$~2.5~km~s$^{-1}$, which makes it likely a non-member (corroborated by its position on the CMD, see \autoref{color mag}). As previously mentioned, we will not include this star in further discussion to ensure self-consistency.

In the case of Tucana~IV, we re-derived the membership probability adding the information of the radial velocity and the corresponding errors obtained in this work, following the approach explained in detail in \cite{battaglia_gaia_2022}. This is shown in \autoref{proper motion}. The stars that fall in the main peak of the radial velocity distribution have a probability $\sim$99\% to be members. We paid particular attention to 5 stars that have a radial velocity that differs more than 2$\sigma$ from the average: stars 9, 16, 17, 22 and 113. According to \cite{battaglia_gaia_2022} method, these stars have a very low probability of being members. Furthermore, considering in particular their offset in radial velocity, we can conclude that these 5 stars are "unlikely members", as we do not expect all of them to be binaries. However, since the Gaia CMD and astrometry are moderately reasonable (see \autoref{color mag} and \autoref{proper motion}), we still analyzed them in Appendix \ref{stars Tuc}.
The systemic radial velocity of Tucana~IV is $\rm v_{sys}~=$~16.2~$\pm$~0.2~km~s$^{-1}$ and the velocity dispersion is $\sigma$~=~5.1 $\pm$ 1.3~km~s$^{-1}$. 

Finally, by checking the position of the member and potential member stars color-coded by the radial velocity (\autoref{ra dec vrad}), we do not see any signatures of tidal disruption in this UFD.

\section{Chemical abundance analysis}
\label{chem text}
We performed this analysis for RGB star member candidates from Section \ref{vrad text} both for metallicities and [C/Fe] abundances (see Table \ref{tab chem}). We adopted the solar reference values from \cite{asplund_chemical_2009}. 
The chemical abundances of the "unlikely members" of Tucana~IV are listed in \autoref{tab stars TucIV} however, if they are not members, the adopted stellar parameters are probably not reliable.

\subsection{Iron}
\label{iron}
We derive the [Fe/H] ratios using the second and third EWs of the $\ion{Ca}{II}$ triplet lines (8542 $\rm \AA$ and 8662 $\rm \AA$) through the empirical relation of \cite{starkenburg_nir_2010} (their Eq. 5). We use these lines because the S/N ratio is not sufficient to accurately measure the metallicity from the weak Fe lines in these metal-poor stars.
We convert our G Gaia magnitude to V Johnson-Cousin magnitude by using the following formula \citep{CarrascoBellazzini2020}:

\begin{align*}
  \rm  G_0-V_0 = &-0.02704 + 0.01424 \, (BP-RP)_0-0.2156\,(BP-RP)_0^2\\
             &+0.01426\,(BP-RP)_0^3
\end{align*}

\begin{figure}[htbp!]
\hspace{-1em}
\includegraphics[width=1.1\linewidth]{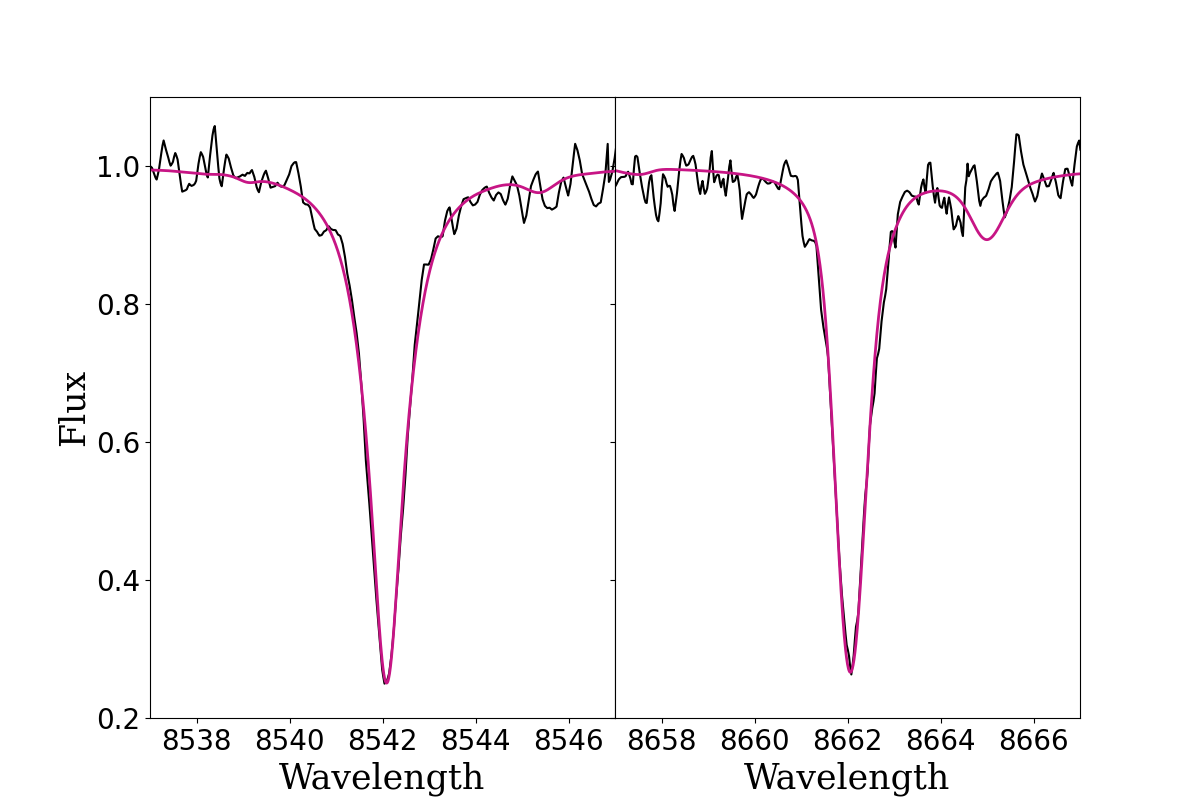} 
\caption{Best fit of EW$_2$ (8542 $\AA$) and EW$_3$ (8662 $\AA$) of star 26 in Grus~II. The black line is the observed spectrum, the magenta line is the synthetic spectrum.}
\label{EW synth}
\end{figure}

The synthetic spectrum of all $\ion{Ca}{II}$ triplet lines was created using Turbospectrum in a well-defined smaller wavelength range (as explained in Section \ref{model}). In this case, it was not possible to gain accurate fits with 1D LTE synthetic spectra with correct parameters due to non-local thermodynamic equilibrium (NLTE) effects, which affect the line profile in a way that is not reproduced with the LTE synthetic spectra, but can be mimicked by changing the input parameters. The EWs were then measured from the best fit synthetic spectrum, minimizing the $\chi^2$ (\autoref{EW synth}).

\begin{figure*}
\hspace{-1em}
\includegraphics[width=0.51\linewidth]{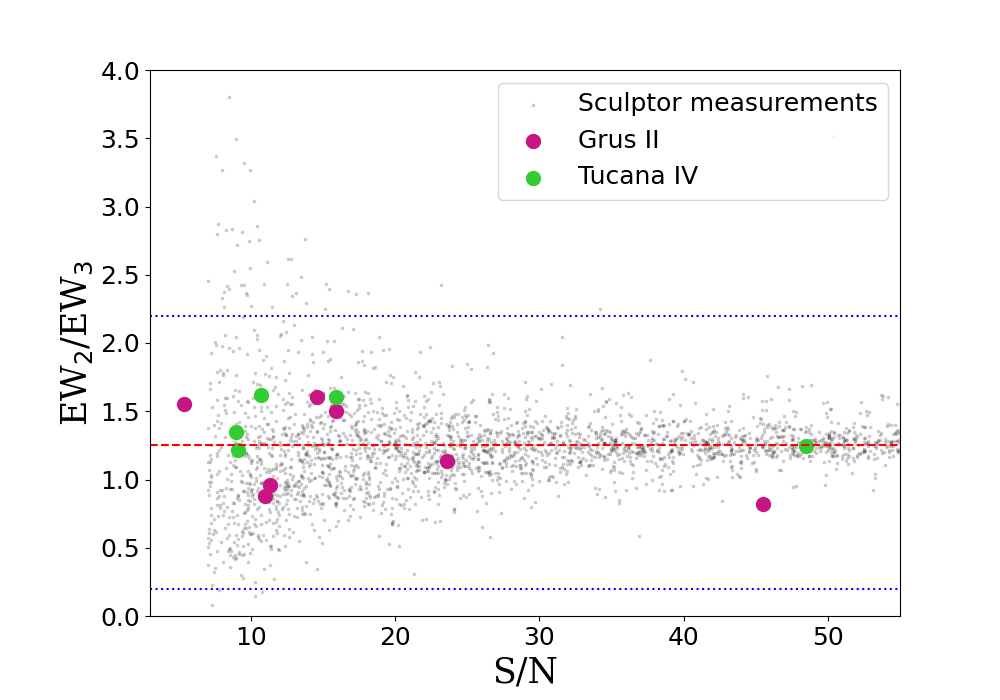}
\hspace{-3em}
\includegraphics[width=0.6\linewidth]{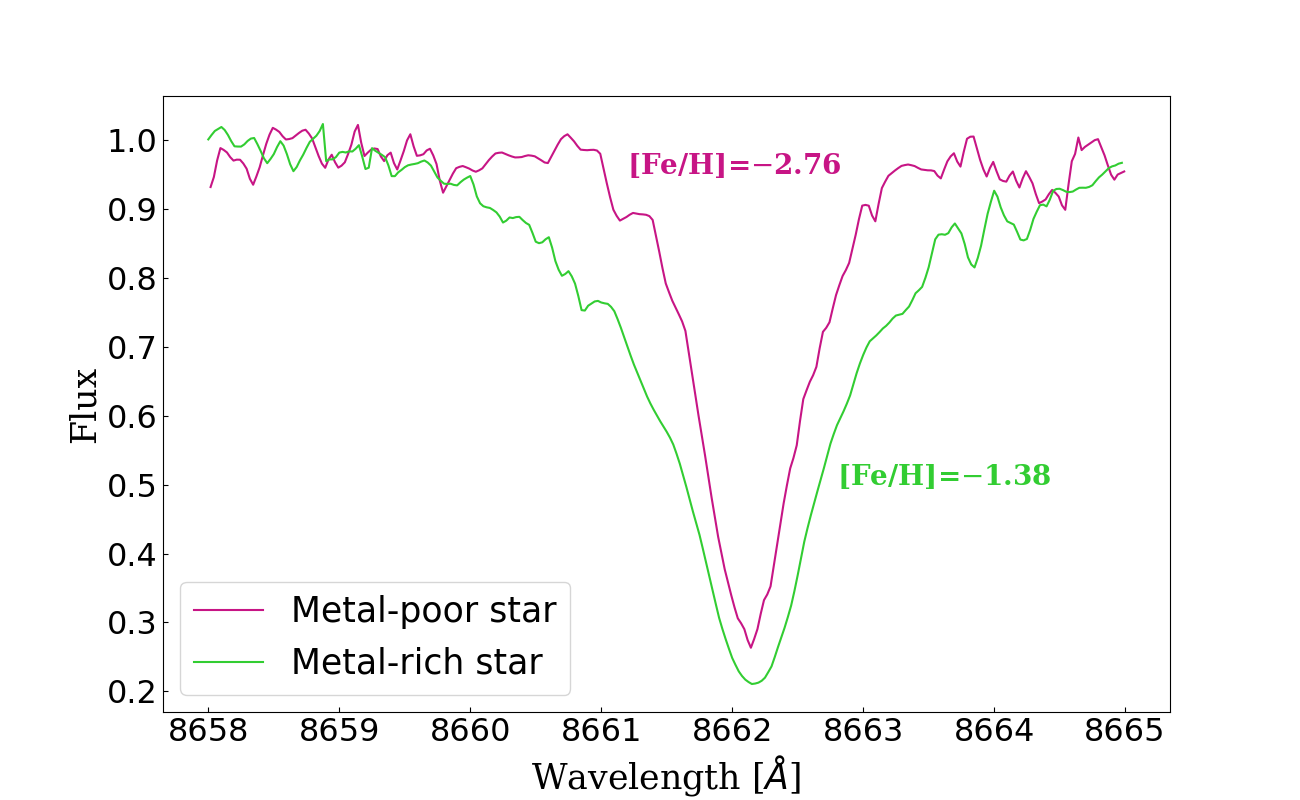}
\caption{\textbf{Left}: Ratios of the EWs of two $\ion{Ca}{II}$ lines (8542 $\angs$ and 8662 $\angs$) as a function of S/N. The two horizontal blue lines are limits of EW2/EW3 at 0.2$<$EW2/EW3$<$2.2, defined as reasonable values by \cite{tolstoy_3d_2023}. \textbf{Right}:
Comparison between spectra of the metal-rich star 5 in Tucana~IV and the metal-poor star 26 in Grus~II.}
\label{EW2/EW3}
\end{figure*}

\cite{tolstoy_3d_2023} reports the ratio between the EWs of the second and third lines of the calcium triplet as a function of S/N for RGB stars in the Sculptor dSph galaxy. To verify that our results are not skewed in the limit of the lowest S/N, \autoref{EW2/EW3} shows the comparison of our measured EW ratio, EW$_2$/\,EW$_3$, with their results. Since we have no measurements in the region beyond the region considered reasonable by \cite{tolstoy_3d_2023}, we infer that the EW ratios are constant with the S/N as expected, indicating that our results are not severely skewed in the limit of low S/N. 
In the right panel of \autoref{EW2/EW3}, we highlight the difference between the reddest $\ion{Ca}{II}$ triplet line (8662 $\angs$) for a metal-poor and a metal-rich star.

\begin{figure*}
\hspace{1em}
\includegraphics[width=0.5\linewidth]{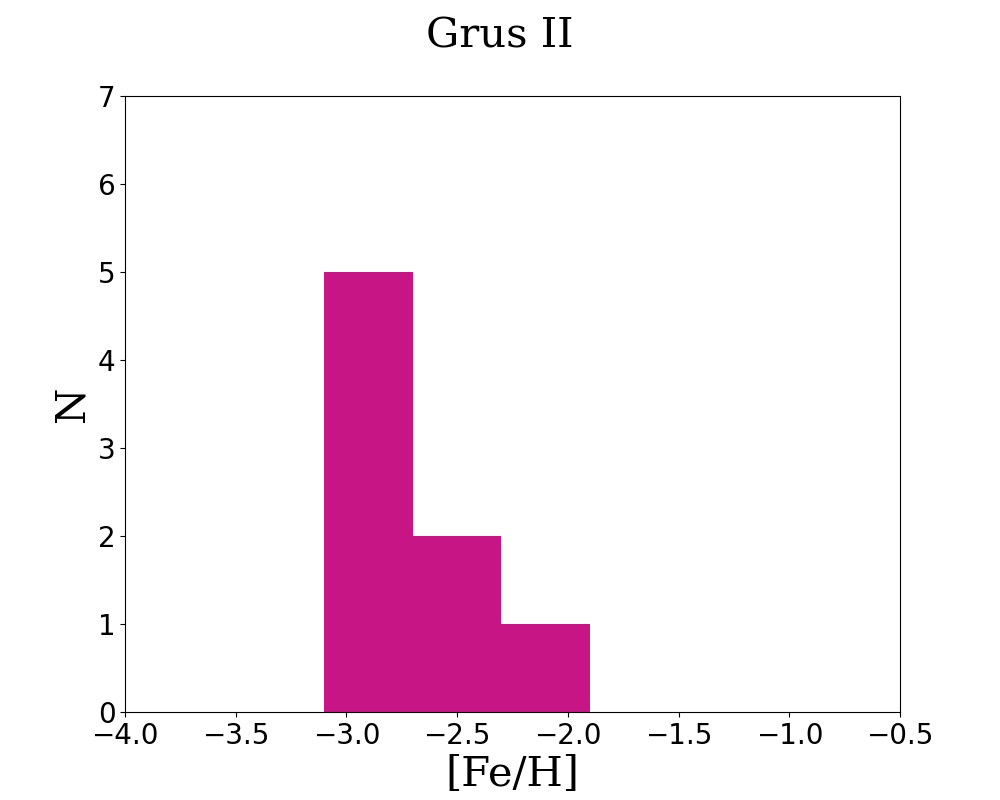} 
\hspace{-1em}
\includegraphics[width=0.5\linewidth]{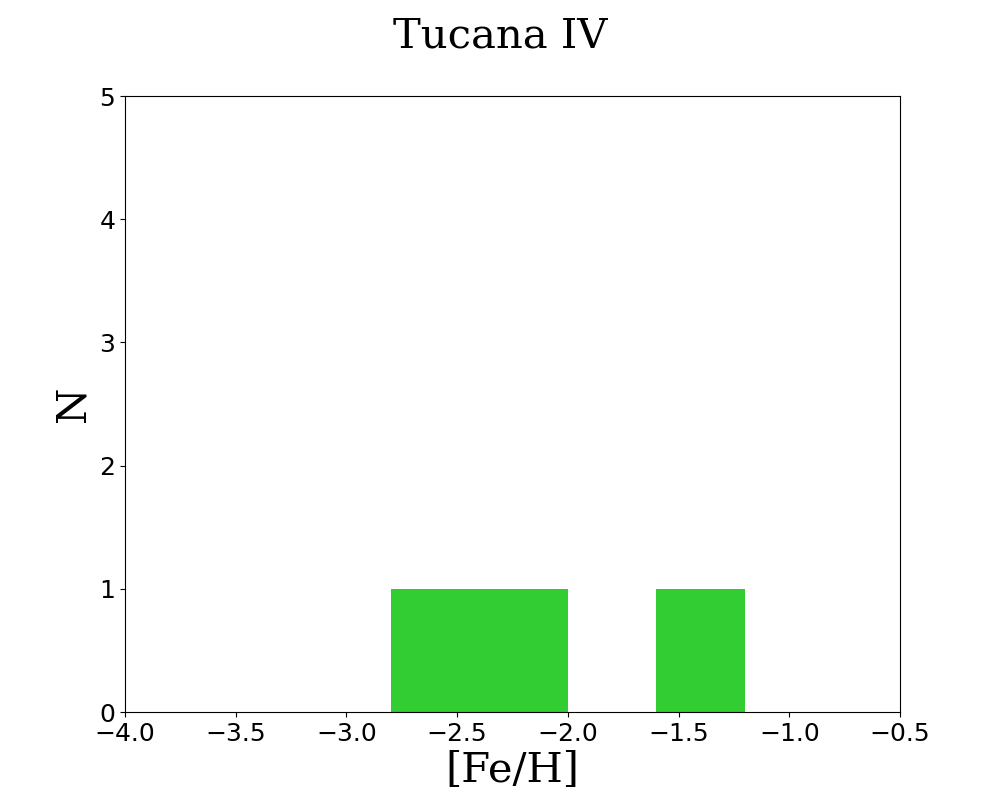}
\caption{Metallicity distribution functions. In Tucana~IV, the "unlikely members" are not included in the distribution. The typical errors on the [Fe/H] are 0.27.}
\label{met dist}
\end{figure*}

\autoref{met dist} shows the Metallicity Distribution Function (MDF) of the member stars in Grus~II and Tucana~IV. The MDF of Grus~II is shifted towards lower metallicity than that of Tucana~IV, leading to a higher probability of finding CEMP-no stars in the first galaxy since these stars are more common at low metallicity \citep[e.g.][]{Beers2005, salvadori_carbon-enhanced_2015}.
Because of the low number statistics (especially in Tucana~IV), we do not draw any concrete conclusion about the star formation history.

According to Starkenburg's formula, the errors on [Fe/H] are obtained with 
\begin{equation}
\sigma_{[\mathrm{Fe/H}]} =
\sqrt{%
\begin{aligned}[t]
& (0.195 + 0.0155 EW_{2+3})^{2}\,\sigma_{V - V_{\mathrm{HB}}}^{2} + (0.458 + \\ &1.3695EW_{2+3}^{-2.5} + 0.0155(V - V_{\mathrm{HB}}))^{2}\,\sigma_{EW_{2+3}}^{2} + \sigma_{\rm par}^2
\end{aligned}%
}
\end{equation}
where $\sigma_{EW_{2+3}}$ is the error on the sum of EW$_{2}$ and EW$_{3}$; $\rm \sigma_{\rm(V-V_{HB})}$ is the error on the mean of the V magnitude of HB stars, ($\rm V_{HB}$=19.14 $\pm$ 0.14 for Grus~II and $\rm V_{HB}$=19.07 $\pm$ 0.14 for Tucana~IV, derived by averaging the V magnitude of the HB stars in our sample); $\sigma_{\rm par}$ is the contribution derived from the fitted parameters, estimated to be 8$\%$ \citep{starkenburg_nir_2010}. The error on the individual EWs is found by adopting the formula from \cite{battaglia2008} (their Eq. 3) and then summed in quadrature. The [Fe/H] values of member stars and their respective errors are reported in \autoref{tab chem}. Moreover, we tested the impact of the systematic error due to the continuum level assuming different S/N, which results $\sigma_{\rm [Fe/H],sys}\lesssim$ 0.05.

We made a check for the brightest stars (5 in Tuc~IV and 26 in Grus II) measuring the [Fe/H] with the Fe lines and the results are in agreement with the Ca triplet measurements.

\subsection{Carbon}

\begin{table}[t]
\caption{Chemical measurements of RGB member stars in Grus II and Tucana IV. [C/Fe] are the carbon abundances corrected using the \cite{placco_carbon-enhanced_2014} tool.}
\label{tab chem}
\renewcommand{\arraystretch}{1.4}
\centering
\setlength{\tabcolsep}{4pt} 
{\small
 \begin{tabular}{ccccccc}
 \multicolumn{6}{c}{\large \textbf{Grus II}}\\
 \hline
 star ID & [Fe/H] & $\sigma_{\rm [Fe/H]}$ & [C/Fe]$_0$ & [C/Fe] & $\sigma_{\rm [C/Fe]}$ \\[0.5ex]
\hline\hline
3   & $-$2.93 & 0.25  & 0.00    & 0.01 & 0.33\\
4   & $-$2.93 & 0.32  & 2.08    & 2.10 & 0.39 \\
9   & $-$2.92 & 0.51 & 1.30    & 1.31 & 0.57 \\
13  & $-$2.69 & 0.26  & 0.76    & 0.77 & 0.33 \\
26  & $-$2.76 & 0.24  & $-$0.34 & $-$0.10 & 0.33 \\
29  & $-$2.24 & 0.27  & 0.98    & 1.00 & 0.35 \\
30  & $-$2.88 & 0.26  & 1.36    & 1.37 & 0.34\\
122 & $-$2.47 & 0.25  & $-$0.78 &$-$0.14 & 0.32\\
\hline
 \end{tabular}
}

\vspace{0.5em}
\setlength{\tabcolsep}{4pt} 
{\small
 \begin{tabular}{ccccccc}
  \multicolumn{6}{c}{\large \textbf{Tucana IV}}\\
\hline
star ID & [Fe/H] & $\sigma_{\rm [Fe/H]}$ & [C/Fe]$_0$ & [C/Fe] & $\sigma_{\rm [C/Fe]}$\\[0.5ex]
\hline\hline
5 & $-$1.38 & 0.28 & $-$0.70 & $-$0.30 & 0.35 \\
7 & $-$2.75 & 0.26 & 0.82 & 0.83 & 0.33 \\
8 & $-$2.09 & 0.27 & 0.04 & 0.05 & 0.36 \\

 \hline
 \end{tabular}
}
\end{table}

\begin{figure}[htbp!]
\includegraphics[width=0.95\linewidth]{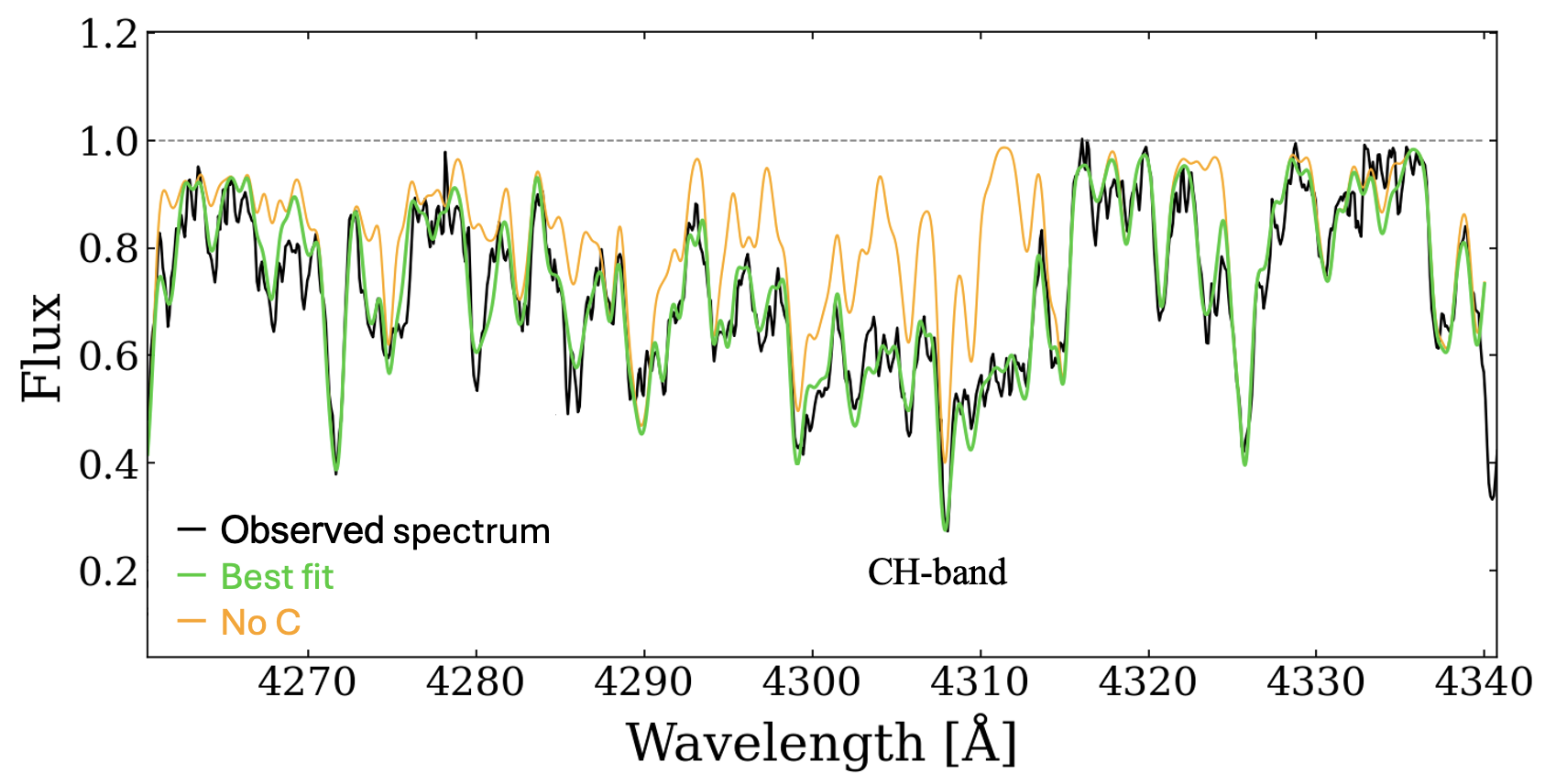} 
\caption{Best fit of the CH band of the star 5 in Tucana~IV. The black line is the observed spectrum, the green line is the synthetic spectrum.}
\label{C synth}
\end{figure}

\label{Carbon text}
To measure the C abundances (\autoref{RGB and HB spectra}), we created a synthetic spectrum as described in Section \ref{vrad text}, adopting the stellar parameters from Section \ref{param} and [Fe/H] from Section \ref{iron}. The synthetic spectrum is generated around the CH-band at 4300 $\angs$ in a wavelength range of 80~$\angs$ (\autoref{C synth}). We used the online tool developed by \cite{placco_carbon-enhanced_2014} to correct the C abundances for the internal mixing occurring in the RGB stars.

\begin{figure*}
\hspace{-1em}
  \includegraphics[width=0.55\linewidth]{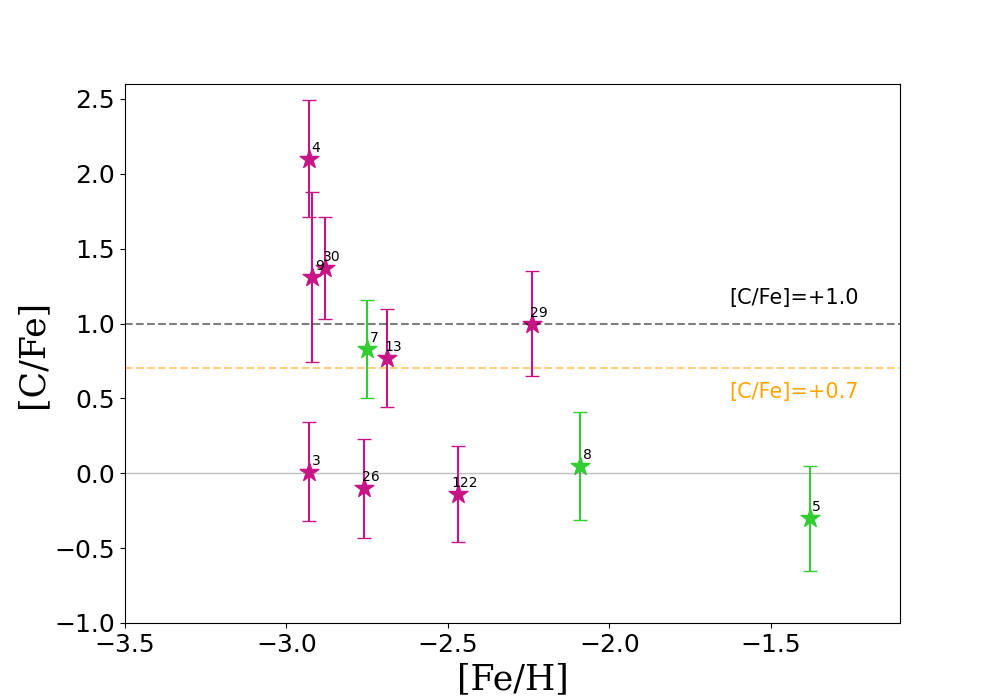}
\hspace{-3em}
  \includegraphics[width=0.55\linewidth]{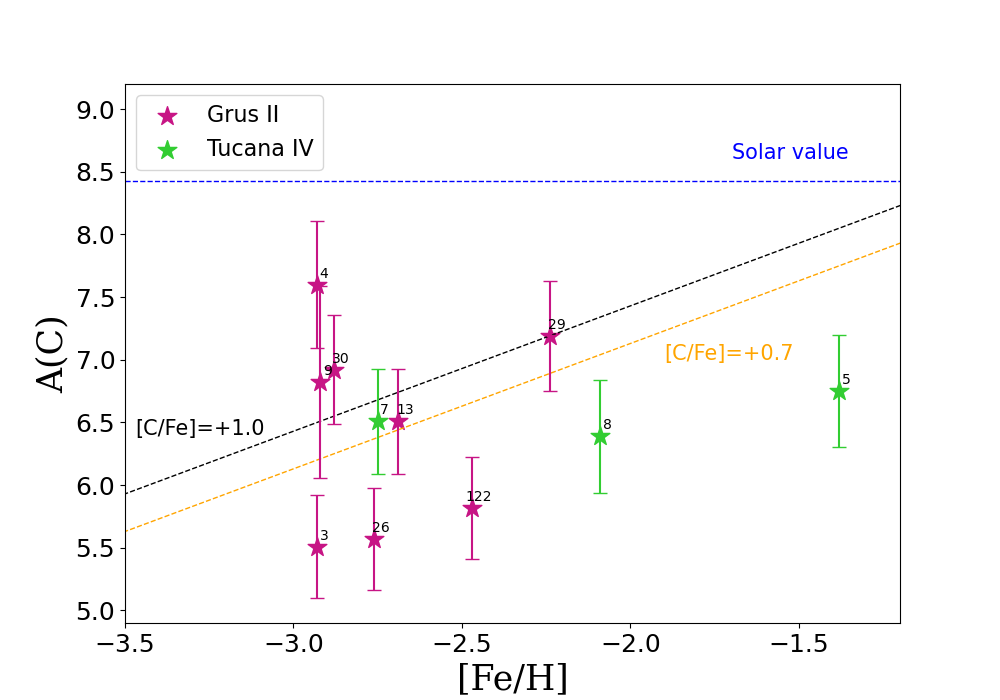}
  
\caption{\textbf{Left}: [C/Fe] as a function of metallicity. The dashed orange line is [C/Fe] = +0.7, while the black dashed line is the other threshold commonly assumed in the literature for CEMP stars, [C/Fe] = +1. \textbf{Right}: The absolute C abundance as a function of  [Fe/H]. The dashed blue line represents the solar value $\rm A(C)_{\odot}$ = 8.43 \citep{asplund_chemical_2009}.}
\label{C_Fe}
\end{figure*}

\autoref{C_Fe} shows the [C/Fe] relation as a function of [Fe/H]. In Grus~II, we identify 3 CEMP stars (stars 4, 9, and 30) with [C/Fe]~$>$~+1 and 2 CEMP stars (stars 13 and 29) with [C/Fe]~$>$~+0.7 and slightly higher [Fe/H]. In Tucana~IV, star 7 has [C/Fe]~$>$~+0.7. All the C measurements are reported in \autoref{tab chem}. 

For these CEMP stars we looked at the \ion{Ba}{II} line at 4554~$\angs$ in LR2. However, since the line is at the edge of the spectrum and S/N is very low (\autoref{tab SN gru}), we were only able to derive upper limits of the Ba abundances:
in Grus~II, stars 4 and 13 have [Ba/Fe] $\lesssim$ 0.6,  star 29 has [Ba/Fe] $\lesssim$ 0.2 and stars 30 and 9 have [Ba/Fe] $\lesssim$ 0.7. In Tucana~IV star 7 has [Ba/Fe] $\lesssim$ 0.2.  
Additionally, we derived the absolute abundances A(C) to verify the nature of these CEMP stars (\autoref{C_Fe}, right): we point out that the stars that fall in the high C-band (i.e. have an absolute C abundance close to the solar value), populate a region which is primarily occupied by CEMP-s stars \citep{bonifacio_topos_2015}. Remembering that CEMP-s stars have [Ba/Fe]~$>$~1 and high absolute C abundances, we can conclude that all the CEMP stars found in Grus~II and star 7 in Tucana~IV are CEMP-no stars. We note that technically these identified CEMP stars could possibly be CEMP-r stars since we do not have Eu measurements. However, this kind of CEMP stars are extremely rare and only a few have been identified in the Milky Way halo.

All the abundances (Fe, C and Ba) were derived under the LTE assumption, allowing us to compare the abundances with other measurements present in the literature. The uncertainties on the C abundances consist of two contributions, which were added quadratically to obtain the total errors. The first is the random error which is found by deriving the [C/Fe] in five different regions of 10~$\angs$ from 4270~$\angs$ to 4320~$\angs$, covering the entire CH-band. We obtained the errors by calculating the standard deviation of these five [C/Fe] measurements. For S/N $>$ 7, $\sigma_{C,rand}$ is $\lesssim$ 0.1 dex, while for 5 $<$ S/N $<$ 7 $\sigma_{C,rand}$ is $\sim$ 0.15 dex\footnote{We remind that the S/N was underestimated in the region of the CH-band because of the lack of continuum points available (see Section \ref{obs}).}. 
The second contribution is due to the stellar parameters and is dominated by the error on the $\teff$, i.e. $\rm \sigma_{C,T_{eff}}\sim$ 0.2 dex. The total error on [C/Fe] was found with the quadratic sum of both the errors on C and Fe abundances (see \autoref{tab chem}).

\section{Comparison with literature}
\label{discussion}

\begin{figure*}
    \includegraphics[width=0.5\linewidth]{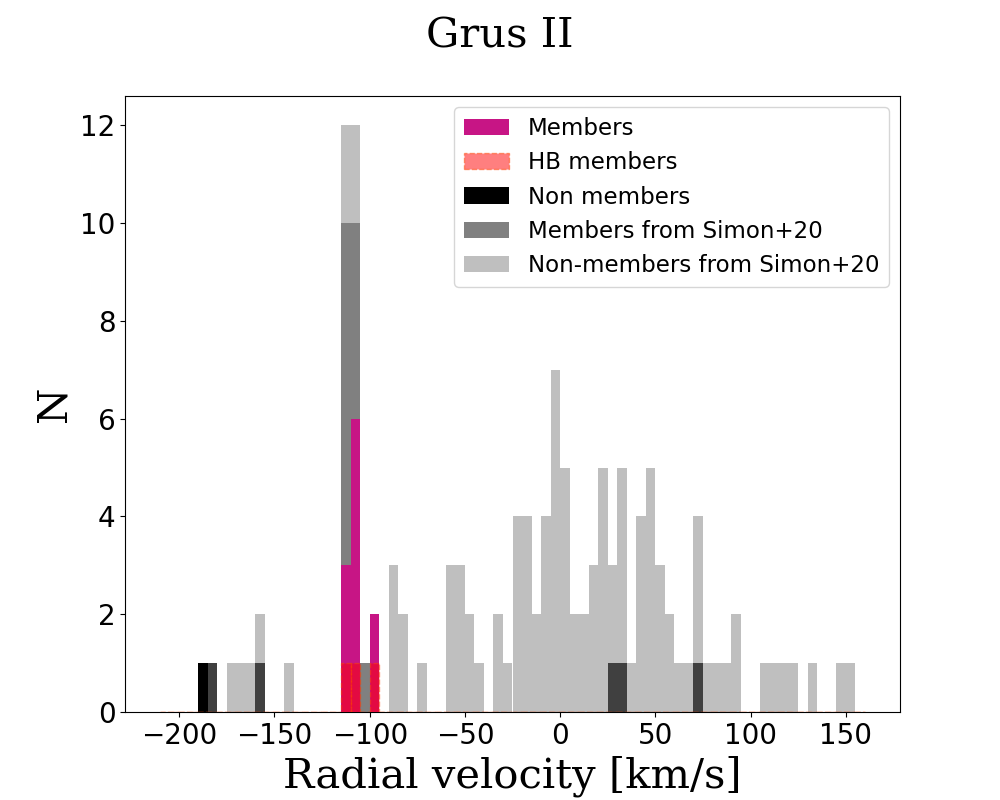}
    \includegraphics[width=0.5\linewidth]{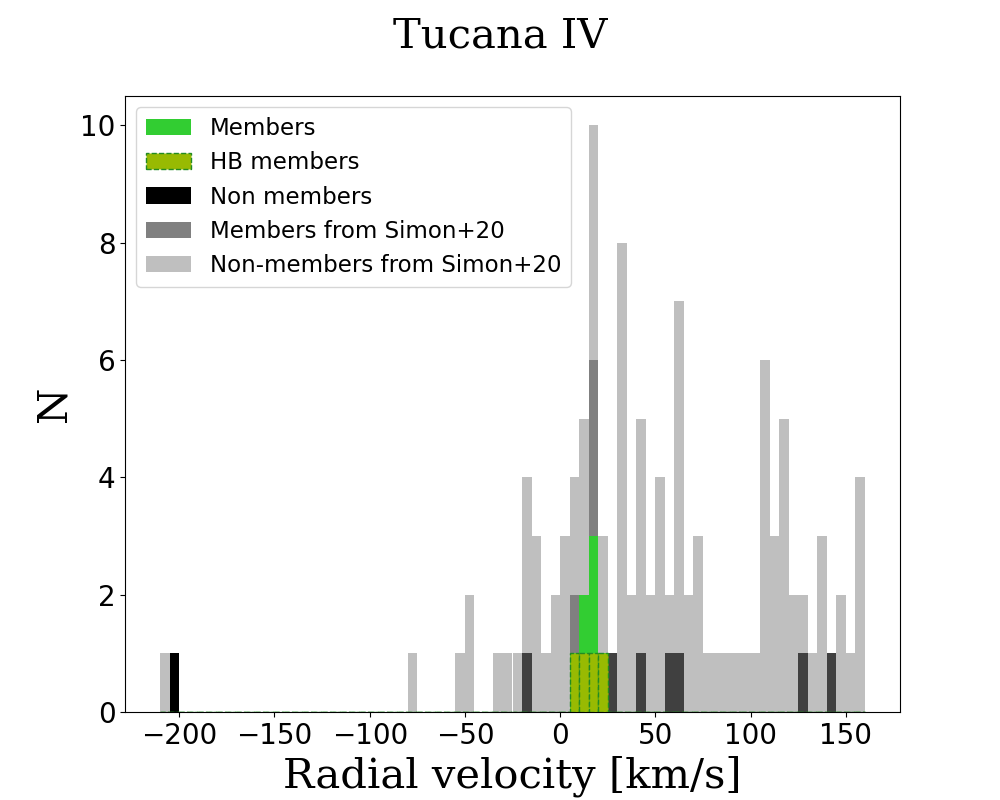}
    \includegraphics[width=0.5\linewidth]{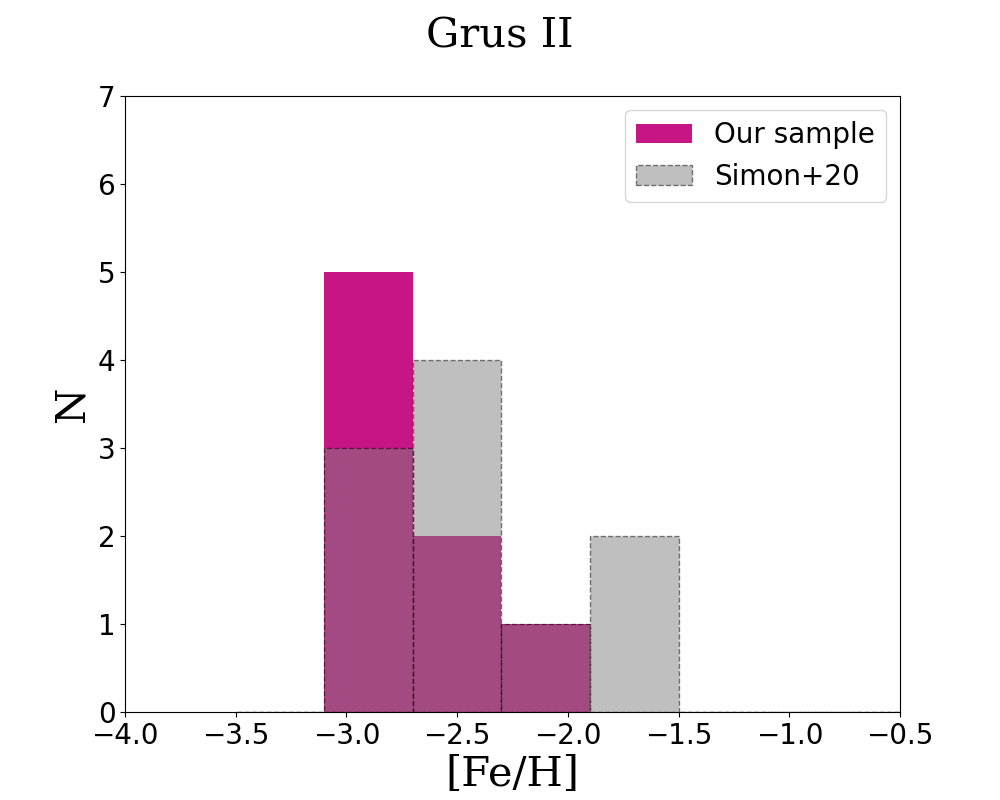}
    \includegraphics[width=0.5\linewidth]{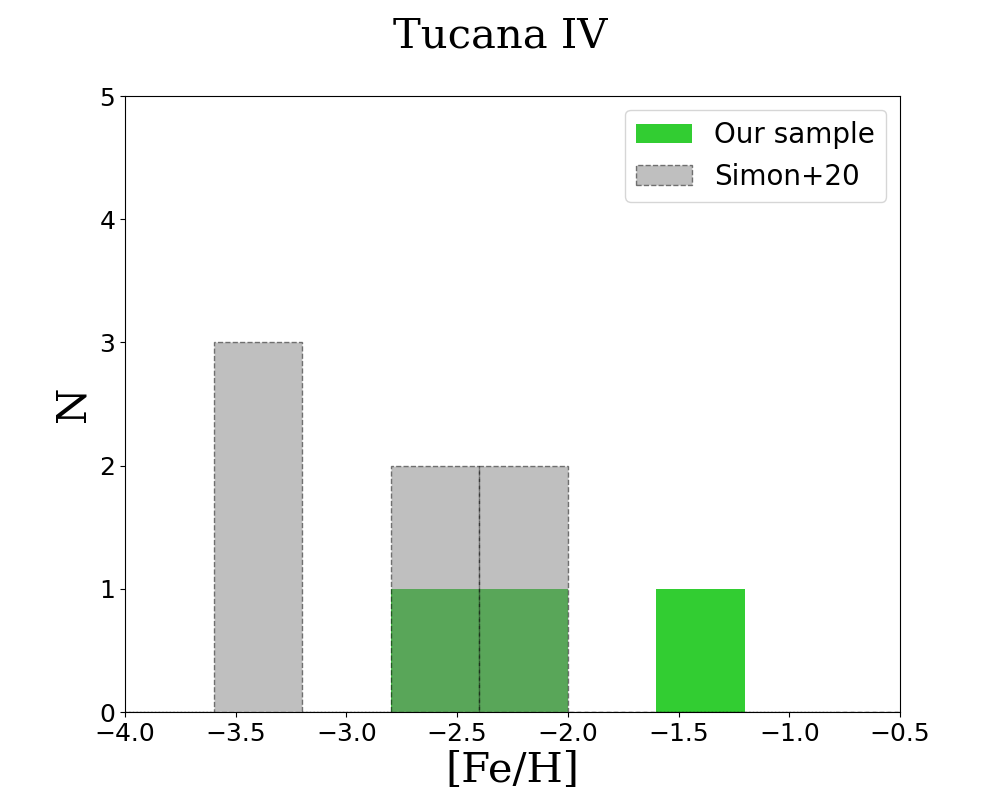}
    \caption{Our radial velocity distributions (upper panels) and metallicity distribution functions (lower panels) compared with \cite{simon_birds_2020}.}
    \label{comp Simon}
\end{figure*} 

Both Grus~II and Tucana~IV have also been investigated by \cite{simon_birds_2020}. They observed stars in Grus~II and Tucana~IV with the IMACS spectrograph on the Magellan/Baade Telescope. Their spectra have R=11~000, in a wavelength range from 7500 to 8900 $\AA$. We checked the presence of stars in common between the two samples: there are 14 stars in Grus~II and 9 stars in Tucana~IV in common, allowing us to make a comparison between the results. The upper panels of \autoref{comp Simon} show the comparison between the radial velocity distributions, while the number values are reported in \autoref{tab com Simon grus} and \autoref{tab com Simon tuc}. The darker grey bars are the members identified by \cite{simon_birds_2020}: our distributions are completely in agreement, corroborating our identification of the members. We recall that the systemic radial velocity we derived for Grus~II is $\rm v_{sys}=-106.5~\pm~1.4$~km~s$^{-1}$, resulting in a small offset from the value reported by \cite{simon_birds_2020} , $\rm v_{sys}=-110.0~\pm~0.5$~km~s$^{-1}$. In the case of Tucana~IV, the systemic radial velocity is $\rm v_{sys}$~=~14.9~$\pm$~1.9~km~s$^{-1}$, considering only the certain members, in good agreement with the value from \cite{simon_birds_2020}, $\rm v_{sys}$ = 15.9$^{+1.8}_{-1.7}$~km~s$^{-1}$.

Additionally, we compare the metallicity distribution functions with those of \citep{simon_birds_2020}. The [Fe/H] measurements of stars in common are consistent within the errors, therefore, despite the low number of stars, the two distributions are in agreement (see the lower panels in \autoref{comp Simon}). Numerical values are reported in \autoref{tab com Simon grus} and \autoref{tab com Simon tuc}.

Finally, \cite{hansen} analyzed the 3 brightest stars in Grus~II (3, 26, 122) from a detailed chemical point of view, providing also the radial velocities (which are in agreement with our results). For the purpose of this work, our attention is focused on the C abundance. All the [Fe/H] measurements and the [C/Fe] of stars 26 and 122 are consistent, while star 3 has quite different [C/Fe], but agrees on being a C-normal star (see Appendix~\ref{app hansen}). Nevertheless, it is not a CEMP star, as we confirm with our abundance derivation.

\section{Discussion and conclusions}
The aim of this work is to analyze the spectra of stars that belong to two ultra-faint dwarf galaxies, Grus~II and Tucana~IV, and derive the carbon abundance in RGB members. In particular, we focused on identifying the CEMP-no stars, whose chemical abundances are consistent with the imprint of primordial low-energy supernovae (E $\lesssim$ $10^{51}$ erg). 

We derived the radial velocity of RGB and HB stars (\autoref{vrad}) using the $\ion{Ca}{II}$ triplet from the FLAMES/Giraffe HR21 spectra and hence determined the membership of the stars to the galaxies by combining our results with Gaia data. Moreover, we compared our results by checking the membership probability reported in \cite{battaglia_gaia_2022} based on Gaia eDR3 (\autoref{proper motion}), where all of our identified members of Grus~II have a membership probability higher than 80$\%$.
We note that 2 of the 13 Grus~II members (stars 1 and 14) are HB stars with only LR2 spectra available, therefore they are not present in the figures and the consequent analysis. In Tucana~IV there are 7 members with a probability over 80$\%$. Afterward, we derived the metallicity distribution functions of the galaxies after an accurate determination of the EWs of the $\ion{Ca}{II}$ triplet. Tucana~IV is shifted to higher metallicity than Grus~II  (\autoref{met dist}), therefore the probability of finding CEMP stars is lower.

Finally, we derived the [C/Fe] to identify carbon-rich stars (\autoref{C_Fe}): 5 CEMP-no stars (thereof 3 with [C/Fe] $>$ +1) were found in Grus~II, while in Tucana~IV we discovered one CEMP-no star. The categorization into CEMP-no stars was based on A(C) and upper limits on Ba (Section \ref{Carbon text}), but further investigation is needed to obtain reliable Ba measurements. Among our identified members in Grus~II and Tucana~IV, 10 stars are very metal-poor ([Fe/H]<$-$2), thereof 6 CEMP-no stars with [C/Fe] $>$ +0.7. According to \cite{2020Ji}, the fraction of CEMP stars at such metallicity is 21 $\pm$ 5\%. In our case this fraction is 60$^{+36}_{-24}\%$ \citep{gehrels_1986}, therefore in these two UFDs the fraction of CEMP-no stars is higher than the average. Considering the threshold of [C/Fe]~$>$~+1, we found 3 CEMP-no stars, therefore 30$^{+29}_{-16}\%$ \citep{gehrels_1986} which is in agreement with \cite{Ji2020} results, i.e. $\approx$ 10 $\pm$ 5\% (assuming similar errors as for [C/Fe] $>$ +0.7).
We can conclude that in Grus~II and Tucana~IV, faint Pop~III SNe likely exploded, enhancing the ISM with high [C/Fe] and maintaining this peculiar abundance ratio in the photosphere of the second-generation stars. However, in the case of Grus~II (where we found 5 CEMP-no stars), we cannot conclude if these stars derived from the same first faint SN or if several such events occur even in this small galaxy (M$_*$=3.4$_{-0.4}^{+0.3}\times 10^3\,\rm M_{\odot}$). Combining this data analysis with literature data could be potentially useful to constrain and understand the star formation history in these small systems.

Even though the UFDs are among the nearest galaxies where it is possible to detect single stars, obtaining spectra of stars in these systems is challenging. This is why there are very few chemical measurements of the populations of these systems in the literature, contained in two principal catalogs, SAGA \citep{saga} and JINA \citep{jinabase}. In the SAGA database which contains the most complete compilation of literature C measurements in metal-poor stars, there are only 78 measurements of carbon abundances in UFDs. By only analyzing a few RGB stars in Grus~II and Tucana~IV, we improve this number by $\sim$14$\%$. This allows us to better understand the CEMP fraction in the UFDs and the imprint that the first stars had in such pristine and ancient systems.

\begin{acknowledgements}\\
This project has received funding from the European Research Council (ERC) under the European Union’s Horizon 2020 research and innovation programme (grant agreement No. 101117455). J. M. Arroyo acknowledges support from the Agencia Estatal de Investigación del Ministerio de Ciencia en Innovación (AEI-MICIN) and the European Social Fund (ESF+) under grant PRE2021-100638. J. M. Arroyo and G. Battaglia acknowledge support from the Agencia Estatal de Investigación del Ministerio de Ciencia, Innovación
y Universidades (MCIU/AEI) under grant “En La Frontera De La Arqueología Galáctica: Evolución De La Materia Luminosa Y Oscura De La Vía Láctea Y Las Galaxias Enanas Del Grupo Local En La Era De Gaia. (FOGALERA)” and the European Regional Development Fund (ERDF) with
reference PID2023-150319NB-C21/10.13039/501100011033.
A.Mucciarelli, acknowledges support from the project "LEGO – Reconstructing the building blocks of the Galaxy by chemical tagging" (P.I. A. Mucciarelli) granted by the Italian MUR through contract PRIN 2022LLP8TK\_001. D.Massari acknowledges financial support from PRIN-MIUR-22: CHRONOS: adjusting the clock(s) to unveil the CHRONO-chemo-dynamical Structure of the Galaxy” (PI: S. Cassisi), and from the INAF Mini Grant 2023 (Ob.Fu. 1.05.23.04.02 – CUP C33C23000960005) CHAM – Chemo-dynamics of the Accreted Halo of the Milky Way (P.I.: M. Bellazzini).
We thank J. Simon for useful discussion and insightful suggestions.
 \end{acknowledgements}

\bibliography{biblio}

\newpage
\onecolumn
\centering
\newpage
\appendix

\section{Isochrones}
\label{Isochrones}
\begin{figure}[H]
    \hspace{-2em}
    \includegraphics[width=0.55\linewidth]{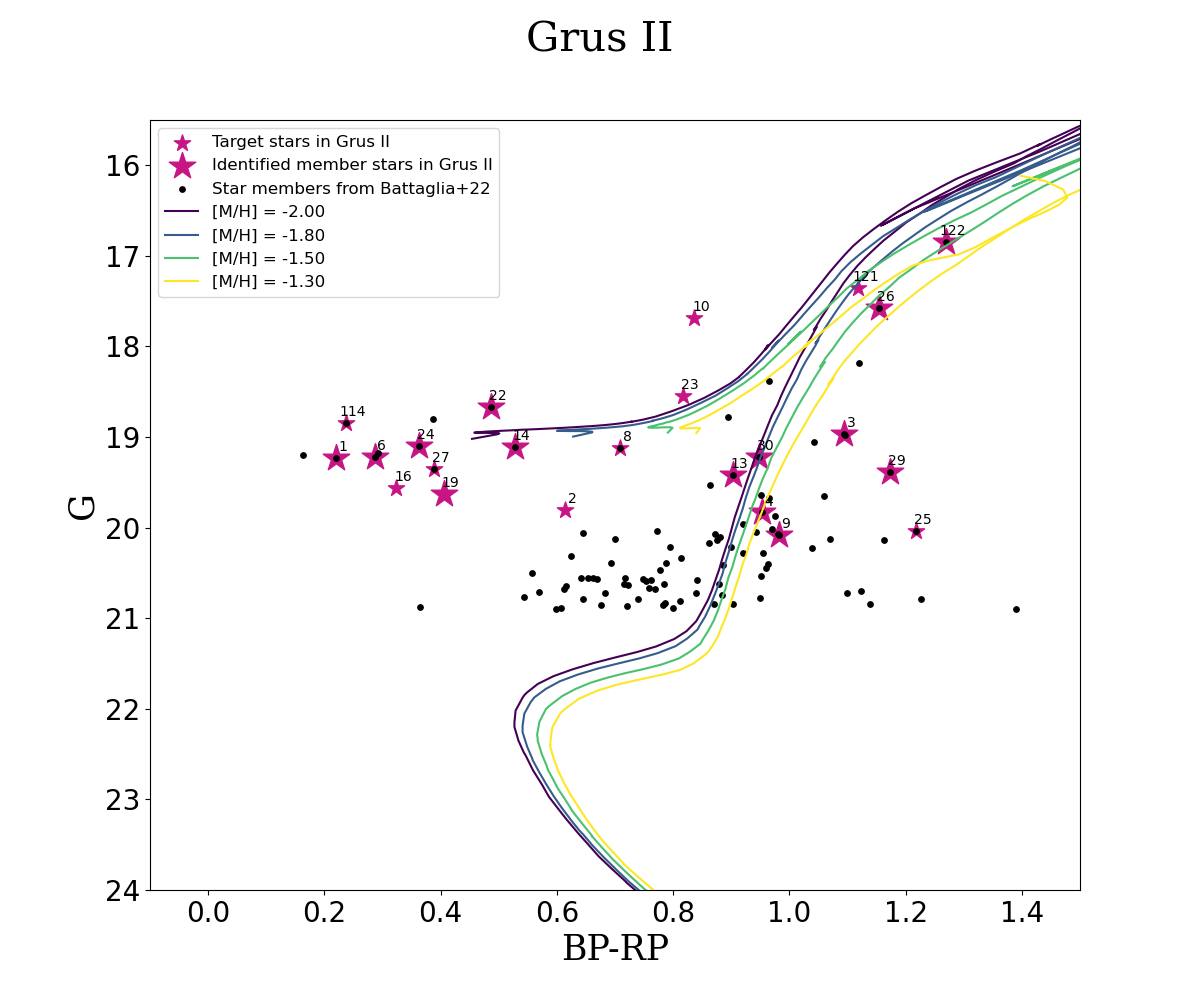}
    \hspace{-3em}
    \includegraphics[width=0.55\linewidth]{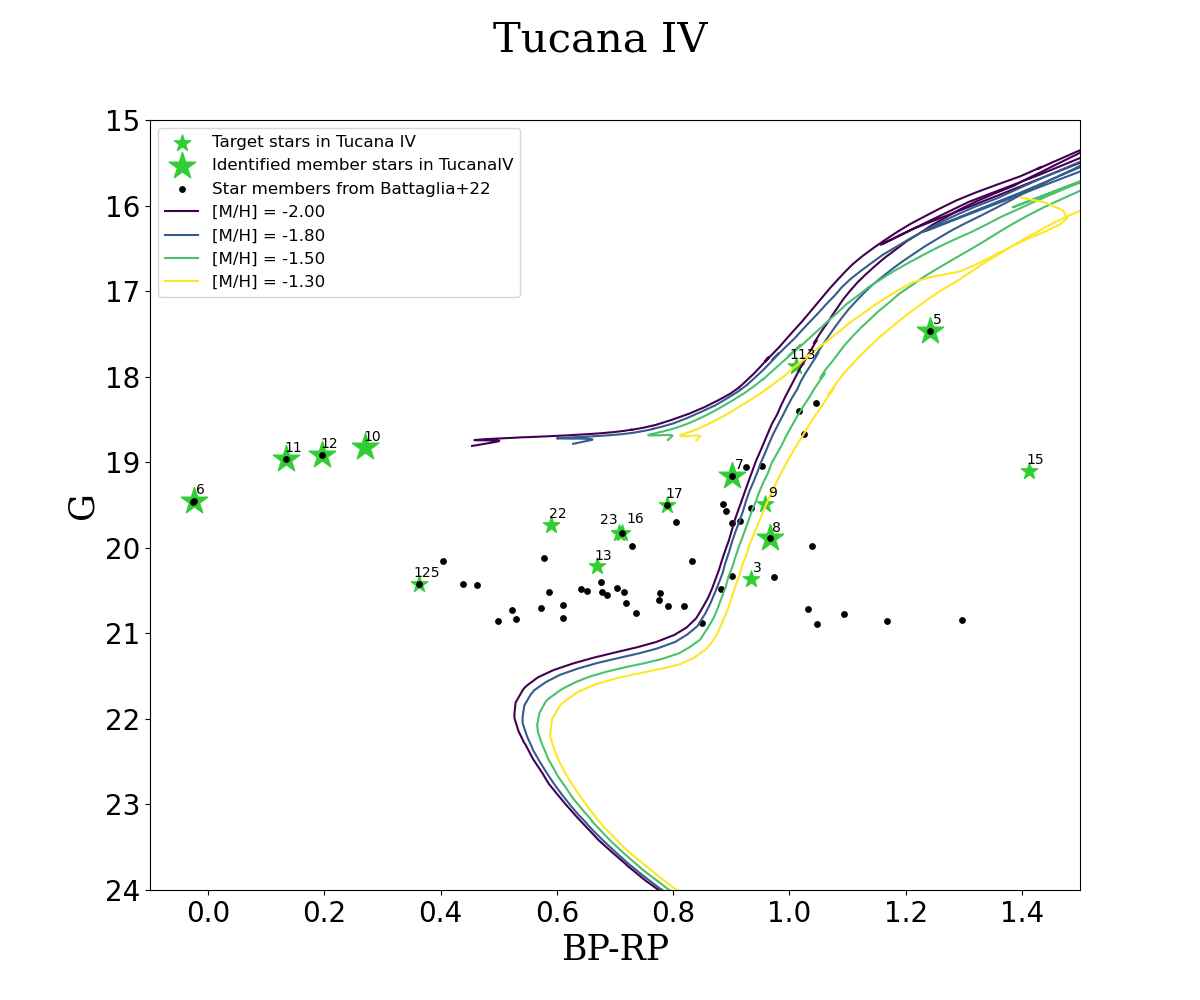}
    \caption{Color-magnitude diagram with isochrones. Our stars fall on the RGB and HB.}
    \label{isochrones}
\end{figure}

\justify
\autoref{isochrones} shows the CMD with isochrones from PARSEC tool \citep{Bressan2012} plotted for different metallicity. We note that below G$<$20 we reach the limit of Gaia, so a spread in photometry is expected because of large uncertainties and our data are not deep enough to reach the main sequence turn-off. The HB of our data agrees with the isochrones, showing that we are in agreement with the distances estimated in \cite{drlica-wagner_eight_2015}.

\section{Radial velocity analysis of near-peak stars in Tucana~IV}
\label{stars Tuc}

\centering
\renewcommand{\arraystretch}{1.4}
\setlength{\tabcolsep}{4pt} 
\begin{table}[H]

\caption{Radial velocity and chemical abundances of the stars 9, 16, 17, 22 and 113 in Tucana~IV.}
\label{tab stars TucIV}
\centering
\begin{tabular}{cccccccc}
\multicolumn{8}{c}{\large \textbf{Tucana IV}}\\
\hline
star ID & v$_{\rm rad}$ & e$_{\vrad}$ & [Fe/H] & $\sigma_{\rm [Fe/H]}$ & [C/Fe]$_0$ & [C/Fe] & $\sigma_{\rm [C/Fe]}$\\
&  [km~s$^{-1}$] &  [km~s$^{-1}$] &  & & & & \\
\hline\hline
9   &  29.3  & 1.3 & $-$1.28 & 0.29 & $-$0.12 & $-$0.10 & 0.36 \\
16  & -18.1  & 1.7 & $-$2.46 & 0.28 & 0.68 & 0.69 & 0.59 \\
17  &  59.3  & 1.7 & $-$2.32 & 0.29 & 0.83 & 1.00 & 0.38 \\
22  &  60.3  & 1.7 & $-$1.86 & 0.28 & 1.70 & 1.71 & 0.38 \\
113 &  43.5  & 0.4 & $-$1.48 & 0.28 & $-$0.05 & 0.05 & 0.35 \\
\hline
\end{tabular}

\end{table}

\justify

Stars 9, 16, 17, 22 and 113 in Tucana~IV are classified as "unlikely members" according to radial velocity determination (Section \ref{vrad text}) and Gaia parameters (proper motions, parallax and position on CMD). However, we performed the chemical abundance analysis described in Section \ref{chem text} also for these stars and the results are reported in Table \ref{tab stars TucIV}.

We derived different radial velocities of star 16 from exposures obtained in different epochs ($-$19.2 km~s$^{-1}$ and $-$16.7 km~s$^{-1}$), which suggests a possible binary star. However, the deviation from the mean velocity of the galaxy is large so its semi-amplitude would be larger than that of binary systems already identified in dwarf galaxies \citep{Koch2014, Hansen2024}, making this star an unlikely member of Tucana~IV. 
Star 22 is a possible CEMP-s star (i.e. likely a binary star, see Section \ref{Carbon text}) which can explain its deviation from the central peak. However, without certain [Ba/Fe] abundance measurements, we cannot confirm its nature. Moreover, the semi-amplitude would be too large for the star to be a Tucana~IV member (as in the case of star 16).
Stars 17 and 113 have acceptable proper motions and parallax, but we do not see any evidence of binarity due in part to having only two epochs. Star 9 differs 5$\sigma_{pm}$ from the systemic proper motion of Tucana~IV (see \autoref{proper motion}). Here, $\sigma_{pm}$ denotes the total dispersion of the proper motions distribution.

Stars 16, 17 and 22 have [C/Fe] consistent with the threshold for CEMP classification. In particular, star 22 has [Ba/Fe] $\lesssim$ 1.5 and falls in the high-C band, therefore could be classified as a possible CEMP-s star. However, we note that should any of these stars be non-members at very different distances compared to the galaxies, the derived stellar abundances, and hence also their chemical abundance measurements, cannot be considered reliable.

\section{Radial velocity in blue}
\label{app vrad blue}
\begin{figure}[H]
    \hspace{-1.5em}
    \includegraphics[width=0.55\linewidth]{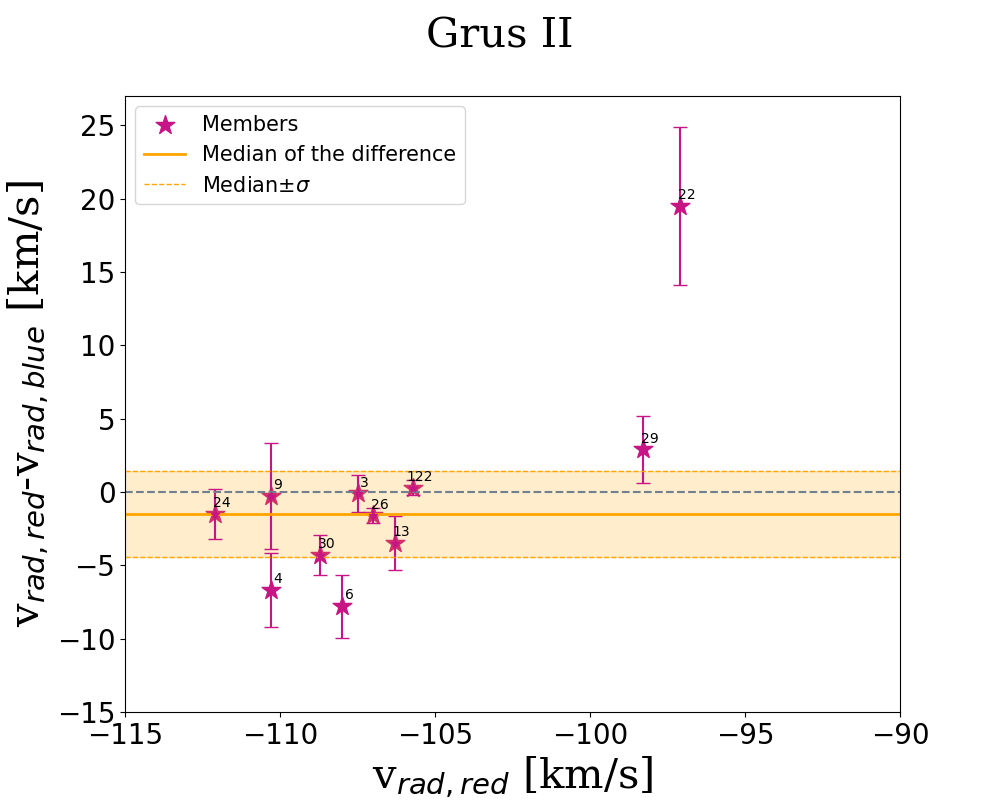}
    \hspace{-1em}
    \includegraphics[width=0.55\linewidth]{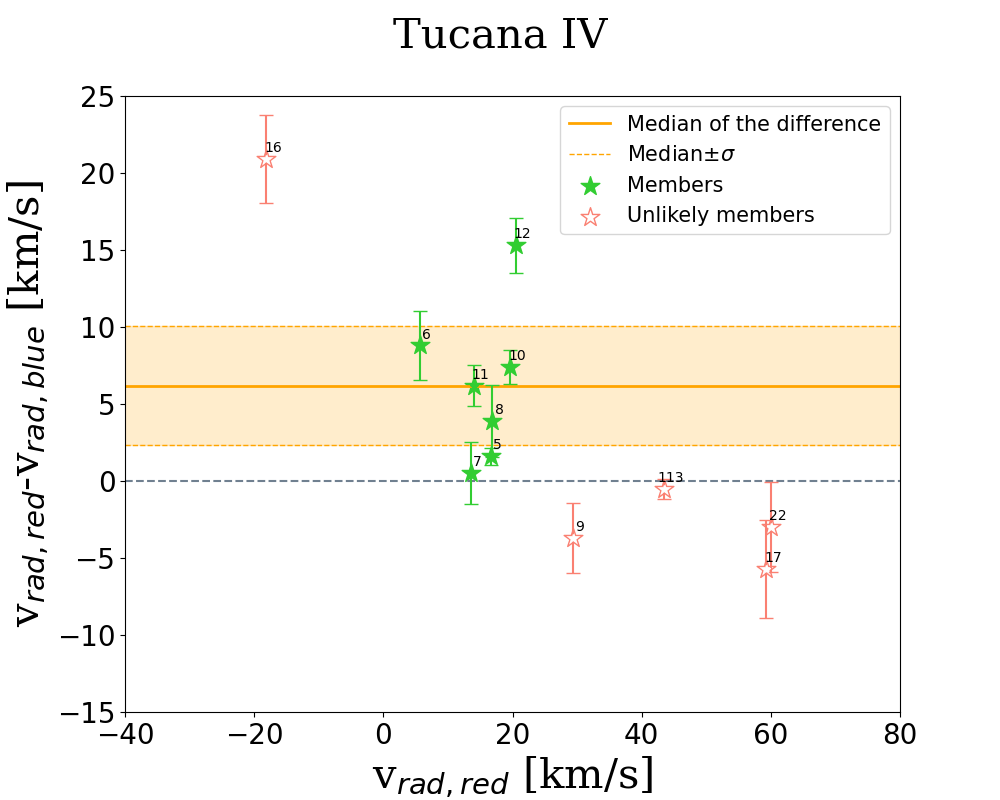}
    \caption{Difference of radial velocity measurements in the red and the blue. The solid line is the mean of the difference, the shaded areas cover $\pm 1\sigma$ from the mean. \textbf{Left:} star 22 shows a big offset which can indicate a possible binary nature and/or be a sign of pulsations. \textbf{Right:} stars 9, 16, 17, 22 and 113 are classified as "unlikely members" as explained in Section \ref{prob}.}
    \label{diff red blue}
\end{figure}

\justify
\autoref{diff red blue} shows the difference between the radial velocities measured from the HR21 and LR2 spectra. The solid line is the mean of the difference, while the dashed lines are $\pm1\sigma$ from the mean. We note that the zero-point in the blue could not be verified because of the lack of skylines, therefore the errors do not include a possible systematic offset. Stars 22 in Grus~II and 12 in Tucana~IV are HB stars and have very different radial velocities between red and blue. However, we remember that the blue spectra present very broad molecular features, lower S/N and resolution that make the radial velocity measurements less trustworthy than the ones from the $\ion{Ca}{II}$ triplet (see Section \ref{vrad text}). 
For these stars, we also obtained the radial velocities from the two epochs available: for star 22 in Grus~II we found $\Delta$v =10~km~s$^{-1}$ and for star 12 in Tucana~IV $\Delta$v=2.5 km~s$^{-1}$, indicating that the hypothesis of binaries could not be excluded. In particular, star 22 in Grus~II is a special case, since the Ca triplet lines are completely contaminated by the skylines as we explained in Section \ref{vrad text}. Moreover, we note that this star corresponds to the variable star V2 studied in \cite{Martinez2019}, who conclude that this is more likely to be a member of the Orphan-Chenab stellar stream. Thus, the radial velocity variation detected in star 22 in our Grus~II sample is likely caused by pulsations.

\section{Comparison with literature results}
\label{app hansen}

In Table \ref{tab com Hansen} are listed the three stars in Grus~II in common with \cite{hansen}. The radial velocities are in agreement within the errors for all the stars. Moreover, the metallicities and carbon abundances are consistent with \cite{hansen} within the errors.

\begin{table}[H]
\caption{Comparison with \citep{hansen} for the three stars in common. The [C/Fe] and A(C) values reported are not corrected for the internal mixing.}
\label{tab com Hansen}
\renewcommand{\arraystretch}{1.3}

\centering
\footnotesize
\setlength{\tabcolsep}{4pt} 
\begin{tabular}{c c c c c c c c c c c c c}
\hline
 star ID  & $\rm T_{eff}$ & $\rm T_{eff, Hansen}$ &  \rm $\rm {log \,\, g}_*$ & ${\rm log \,\, g}_{\rm *, Hansen}$ & $\vrad$  & v$_{\rm rad,Hansen}$ & [Fe/H] & [Fe/H]$_{\rm Hansen}$ & [C/Fe] & [C/Fe]$_{\rm Hansen}$ & A(C) & A(C)$_{\rm Hansen}$\\
 & [K] & [K] & &  &[km~s$^{-1}$] & [km~s$^{-1}$] & & & & & &\\
\hline\hline
3 & 4944 & 5121 & 2.27 & 1.91 &$-$107.5 & $-$106.5 & $-$2.93 & $-$2.94 & 0.00 & +0.52 & 5.50 & 6.01\\
26 & 4823 & 4740 & 1.66 & 1.55 &$-$107.0 & $-$106.6 & $-$2.76 & $-$2.69 & $-$0.34 & $-$0.33 & 5.33 & 5.41\\
122 & 4614 & 4556 & 1.27 & 1.22 & $-$105.7 & $-$106.9 & $-$2.47 & $-$2.49 & $-$0.78 & $-$0.72 & 5.18 & 5.22\\
\hline
\end{tabular}
\end{table}

\begin{figure}[H]
 \hspace{-2em}
    \includegraphics[width=0.6\linewidth]{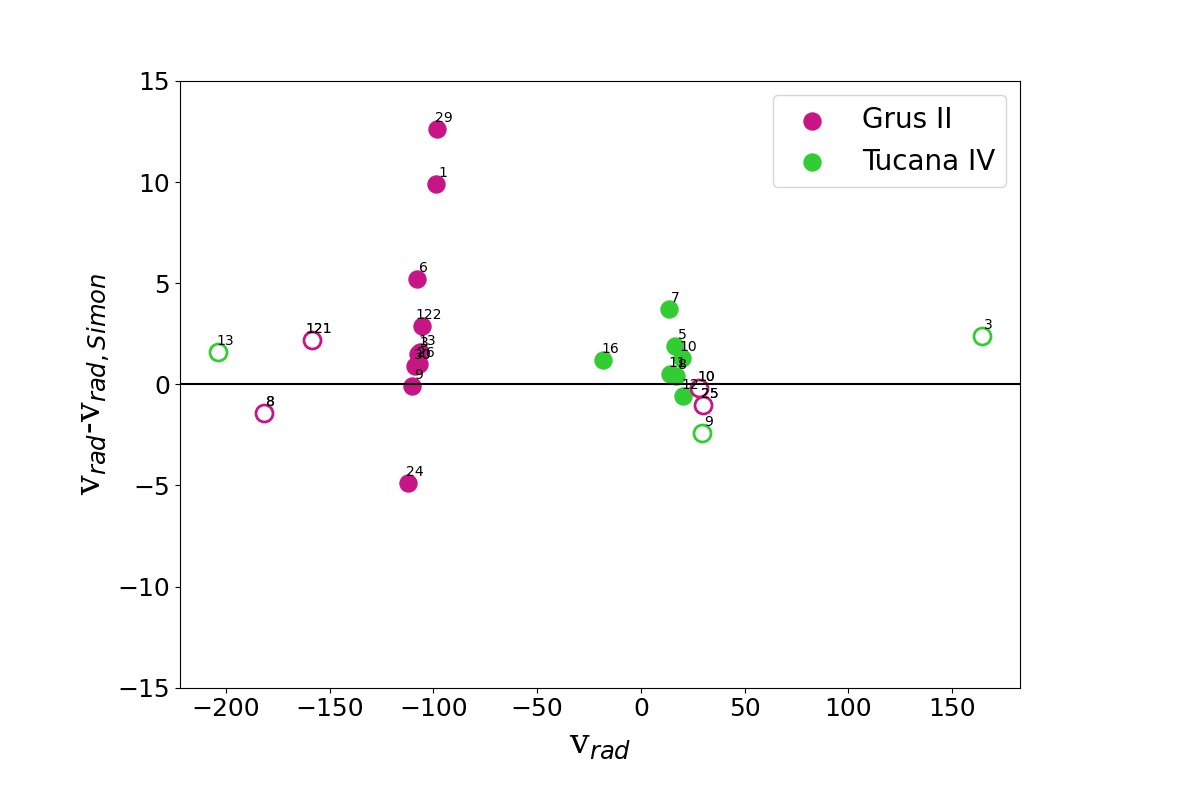}
    \hspace{-4em}
    \includegraphics[width=0.6\linewidth]{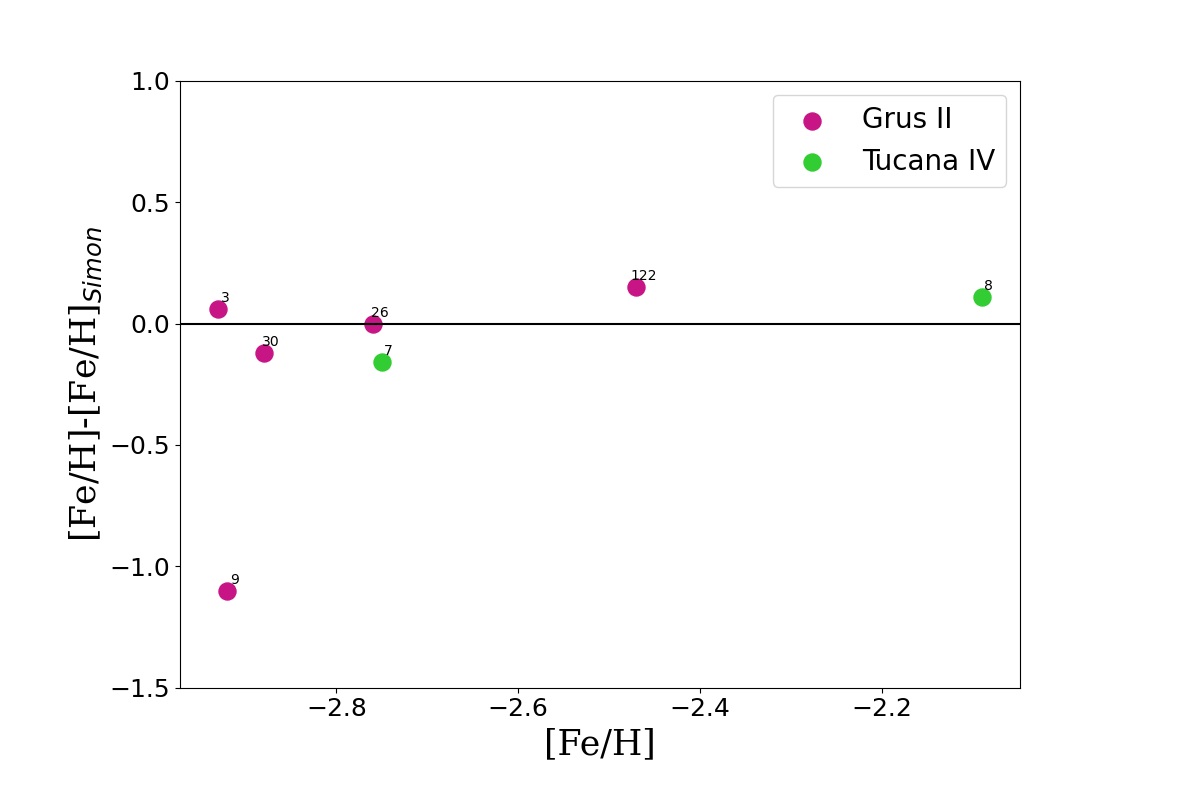}
    \caption{Comparison between our and \cite{simon_birds_2020} results in radial velocity (left) and [Fe/H] (right). The empty symbols are the non-member stars.}
    \label{fig Fe hansen}
\end{figure}

\section{Additional Tables}

\begin{table}[H]
\centering
\caption{The number of exposures in HR21 and LR2 settings with the corresponding S/N per pixel values, and radial velocities from the Ca triplet with corresponding uncertainties. The italic ID identifies the stars whose radial velocities are derived from the blue spectra, while the "*" symbol represents the HB stars. We note that stars 14 and 22 in Grus~II correspond to variable stars V1 and V2 studied in \cite{Martinez2019}. They concluded that star 14 (their V1) is a Grus~II member, but HB star 22 (their V2) is more likely to be a member of the Orphan-Chenab stellar stream.}
\begin{minipage}{0.48\textwidth}
\centering

\label{tab SN gru}
\renewcommand{\arraystretch}{1.3}
\centering
\footnotesize
\setlength{\tabcolsep}{4pt} 
\begin{tabular}{c c c c c S[table-format=3.1] c}
\multicolumn{7}{c}{\large \textbf{Grus II}}\\
\hline
 star ID & HR21 & LR2 &  [S/N]$_{\rm HR21}$ & [S/N]$_{\rm LR2}$ & v$_{\rm rad}$ &  $\rm e_{\vrad}$\\
 & & & pix$^{-1}$ & pix$^{-1}$ & [km~s$^{-1}$]& [km~s$^{-1}$]\\
\hline\hline
 \textit{1}*            &  $-$    &  2   &   $-$    &   21   &    -98.8      &   0.6  \\ 
 3            &  2     &  4   &   24    &   12   &   -107.5      &   0.5  \\ 
 4            &  2     &  2   &   11    &    7   &   -110.3      &   1.2 \\
 6*           &  2     &  2   &    7    &   20   &   -108.0      &   1.8  \\
 8            &  2     &  2   &   15    &    6   &   -181.9      &   0.8  \\
 9            &  4     &  4   &    5    &    6   &   -110.3      &   2.5  \\
 10           &  2     &  2   &   37    &   14   &     28.2      &   0.3  \\ 
 13           &  2     &  2   &   15    &   10   &   -106.3      &   0.9  \\ 
\textit{14}*   &  $-$    &  2   &   $-$    &   12   &  -104.0    &   1.1  \\ 
 16*          &    4   &  4   &    7    &   11   &   -189.7      &   2.0  \\ 
 22*          &    2   &  2   &   11    &   21   &    -97.1      &   1.1  \\ 
 23           &    2   &  2   &   24    &    4   &     74.9      &   0.5  \\ 
 24*          &    2   &  2   &   10    &   15   &   -112.1      &   1.3  \\ 
 25           &    4   &  4   &   13    &    2   &     30.1      &   1.0  \\ 
 26           &    2   &  2   &   46    &   22   &   -107.0      &   0.3  \\  
 27*          &    2   &  2   &    7    &   13   &    171.1      &   1.7  \\ 
 29           &    2   &  2   &   11    &    8   &    -98.3      &   1.1  \\
 30           &    2   &  2   &   16    &   12   &   -108.7      &   0.8  \\  
 \textit{114}* &    $-$  &  2   &   $-$    &    8   &    -22.4   &   1.6 \\ 
 121          &    2   &  2   &   52    &   26   &   -158.6      &   0.2  \\ 
 122          &    2   &  2   &   71    &   19   &   -105.7      &   0.2 \\ 
\hline            
\end{tabular}
\end{minipage}
\hfill
\begin{minipage}{0.48\textwidth}
\centering
\renewcommand{\arraystretch}{1.3}
\centering
\footnotesize
\setlength{\tabcolsep}{4pt} 
\vspace{-6em}
\begin{tabular}{c c c c c S[table-format=3.1] c }
\multicolumn{7}{c}{\large \textbf{Tucana IV}}\\
\hline
 star ID & HR21 & LR2 &  [S/N]$_{\rm HR21}$ & [S/N]$_{\rm LR2}$ & v$_{\rm rad}$ &  $\rm e_{\vrad}$\\
 & & & pix$^{-1}$ & pix$^{-1}$ & [km~s$^{-1}$]& [km~s$^{-1}$]\\
\hline\hline
 3              &    4   &   4   &    7  &    3   &   164.4     &  1.7    \\ 
 5              &    4   &   4   &   49  &   20   &    16.6     &  0.2    \\
 6*             &    4   &   4   &    7  &   15   &     5.6     &  1.8    \\
 7              &    4   &   4   &   16  &    8   &    13.5     &  0.8    \\
 8              &    4   &   4   &   11  &    8   &    16.9     &  1.2    \\
 9              &    4   &   4   &   11  &    8   &    29.3     &  1.1    \\  
 10*            &    4   &   4   &   13  &   21   &    19.6     &  0.9    \\ 
 11*            &    4   &   4   &   11  &   21   &    14.0     &  1.2    \\ 
 12*            &    4   &   4   &   10  &   13   &    20.5     &  1.3    \\ 
 13             &    4   &   4   &    9  &    5   &  -203.8     &  1.5    \\  
 15             &    4   &   4   &   12  &    2   &   143.3     &  1.0    \\  
 16             &    4   &   4   &    9  &    6   &   -18.1     &  1.4    \\ 
 17             &    4   &   4   &    9  &    5   &    59.3     &  1.4    \\ 
 22             &    4   &   4   &    9  &    6   &    60.3     &  1.4    \\  
 23             &    4   &   4   &    8  &    4   &   125.3     &  1.6    \\  
 113            &    4   &   4   &   26  &   20   &    43.5     &  0.5    \\ 
 \textit{125}*  &    $-$  &   4   &  $-$ &    6   &   -40.0     &  2.2    \\ 
\hline

\end{tabular}
\end{minipage}

\end{table}

\begin{sidewaystable}
\caption{Stellar parameters and Gaia DR3 data for Grus II.}
\label{tab par G}
\renewcommand{\arraystretch}{1.3}
\centering
\footnotesize
\setlength{\tabcolsep}{4pt} 

 \begin{tabular}{c c c c c c c c c c c c c c c c}
\hline
Gaia ID & star ID & RA & DEC & G & BP & RP & (BP $-$ RP)$_0$ &  (BP $-$ G)$_0$ & (G$-$RP)$_0$  & $\mu_{\alpha}$  &  $\mu_{\delta}$ &  T$_{\rm eff}$ & $\sigma_{\rm T_{eff}}$ & $\rm {log \,\, g}_*$ & v$_{\rm turb}$ \\
& & & & & & & & & & [mas/yr] & [mas/yr] & [K] & [K] & & [km~s$^{-1}$]\\
\hline\hline
6561433598368152192 & 1 & 331.08387 & $-$46.39487 &  19.2333 & 19.3114 & 19.0916 &  0.1919 &  0.0679 &   0.1240  & 0.273 & $-$1.624 & >\,6500 & 140 & 3.20 & 1.36\\ 
6561421778618145792 & 2 & 330.92058 & $-$46.40760 &  19.8028 & 20.0295 & 19.4149 &  0.5913 &  0.2166 &   0.3748  & 0.069 & $-$0.051 & 6146 &  80 & 3.04 & 1.39\\ 
6561421645474640000 & 3 & 330.96673 & $-$46.41297 &  18.9611 & 19.4572 & 18.3629 &  1.0722 &  0.4846 &   0.5876  & 0.207 & $-$1.782 & 4944 & 106 & 2.27 & 1.55 \\ 
6561410199386304384 & 4 & 331.01845 & $-$46.42266 &  19.8327 & 20.2353 & 19.2832 &  0.9289 &  0.3910 &   0.5379  & 0.127 & $-$2.107 & 5238 &  87 & 2.75 & 1.45\\
6561432464496743296 & 6 & 331.14065 & $-$46.44445 &  19.2171 & 19.3320 & 19.0454 &  0.2595 &  0.1047 &   0.1548  & 0.493 & $-$1.445 & >\,6500 & 162 & 3.12 & 1.38\\
6561419270357129856 & 8 & 330.88169 & $-$46.53853 &  19.1157 & 19.4518 & 18.7424 &  0.6875 &  0.3262 &   0.3614  & $-$0.155 & $-$1.264 & 5853 & 202 & 2.68 & 1.46\\
6561433804526599168 & 9 & 331.06859 & $-$46.37375 &  20.0840 & 20.5170 & 19.5352 &  0.9578 &  0.4209 &   0.5368  & 0.606 & $-$2.154 & 5178 & 105 & 2.82 & 1.44\\
6561414498649007488 & 10 & 330.80891 & $-$46.51693 & 17.6815 & 18.0132 & 17.1767 &  0.8161 &  0.3221 &   0.4940  &  $-$0.258 & $-$2.036 & 5507 &  80 & 1.99 & 1.60\\  
6561408279535319936 & 13 & 331.07562 & $-$46.54788 & 19.4178 & 19.7541 & 18.8516 &  0.8813 &  0.3260 &   0.5553  & 0.239 & $-$1.310 & 5349 &  94 & 2.63 & 1.47\\ 
6561426485901552640 & 14 & 330.87292 & $-$46.28101 & 19.1054 & 19.3193 & 18.7921 &  0.5029 &  0.2036 &   0.2993  & 0.496 & $-$1.527 & 6430 & 126 & 2.84 & 1.43\\ 
6561434594800593024 & 16 & 331.21856 & $-$46.35556 & 19.5657 & 19.7328 & 19.4089 &  0.3000 &  0.1580 &   0.1420  & $-$0.442 & $-$2.874 & >\,6500 & 285 & 3.22 & 1.36\\ 
6561404633108542976 & 19 & 331.21860 & $-$46.58282 & 19.6297 & 19.8025 & 19.3969 &  0.3802 &  0.1627 &   0.2175  & $-$0.149 & $-$0.166 & 6878 & 172 & 3.17 & 1.37\\  
6561409409112267264 & 22 & 331.02493 & $-$46.48206 & 18.6689 & 18.8711 & 18.3852 &  0.4628 &  0.1926 &   0.2702  & 0.318 & $-$1.702 & 6569 & 147 & 2.71 & 1.46\\  
6561436072268568064 & 23 & 331.10737 & $-$46.25574 & 18.5485 & 18.8842 & 18.0676 &  0.7960 &  0.3259 &   0.4700  & $-$120 & $-$1.028 & 5557 &  92 & 2.35 & 1.53\\  
6561420163710392704 & 24 & 330.92204 & $-$46.46660 & 19.1005 & 19.2505 & 18.8884 &  0.3387 &  0.1409 &   0.1979  & 0.631 & $-$1.467 & >\,6500 & 163 & 3.00 & 1.40\\ 
6561445830435058816 & 25 & 331.03710 & $-$46.32437 & 20.0408 & 20.5159 & 19.2976 &  1.1964 &  0.4632 &   0.7333  & 0.386 & $-$1.749 & 4695 & 116 & 2.59 & 1.48\\ 
6561445628572018944 & 26 & 331.04160 & $-$46.35068 & 17.5749 & 18.0758 & 16.9214 &  1.1316 &  0.4887 &   0.6429  & 0.421 & $-$1.434 & 4823 &  81 & 1.66 & 1.67\\  
6561407278808404096 & 27 & 330.99526 & $-$46.57928 & 19.3498 & 19.4765 & 19.0874 &  0.3651 &  0.1173 &   0.2478  & 0.353 & $-$2.28 & 6968 3 & 83 & 3.08 & 1.38\\  
6561410199386302592 & 29 & 331.00051 & $-$46.42478 & 19.3876 & 19.9808 & 18.8078 &  1.1511 &  0.5814 &   0.5697  & 0.355 & $-$1.380 & 4823 & 178 & 2.39 & 1.52 \\  
6561421572459692416 & 30 & 330.99051 & $-$46.43441 & 19.2148 & 19.6126 & 18.6647 &  0.9256 &  0.3867 &   0.5388  & 0.442 & $-$1.645 & 5246 &  86 & 2.50 & 1.50\\  
6561431949100750592 & 114 & 331.30712 & $-$46.33453 & 18.8423 & 18.9432 & 18.7062 &  0.2155 &  0.0930 &  0.1226  & 0.584 & $-$948 & >\,6500 & 191 & 3.02 & 1.40\\  
6561433052907770752 & 121 & 331.17323 & $-$46.37917 & 17.3589 & 17.8510 & 16.7322 &  1.0960 &  0.4801 &  0.6159  & 0.360 & $-$2.111 & 4892 &  88 & 1.61 & 1.68\\  
6561403567956557312 & 122 & 331.09963 & $-$46.61741 & 16.8500 & 17.4057 & 16.1368 &  1.2486 &  0.5444 &  0.7043  & 0.414 & $-$2.111 & 4614 &  79 & 1.27 & 1.75\\ 
\hline

 \end{tabular}

\end{sidewaystable}

\begin{sidewaystable}
\caption{Stellar parameters and Gaia DR3 data for Tucana IV.}
\label{tab par T}
\renewcommand{\arraystretch}{1.3}
\centering
\footnotesize
\setlength{\tabcolsep}{4pt} 

\begin{tabular}{c c c c c c c c c c c c c c c c}
\hline
Gaia ID & star ID & RA & DEC & G & BP & RP & (BP $-$ RP)$_0$ &  (BP $-$ G)$_0$ & (G$-$RP)$_0$  & $\mu_{\alpha}$ & $\mu_{\delta}$ & T$\rm _{eff}$ &  $\sigma_{\rm T_{eff}}$ & $\rm {log \,\, g}_*$ & v$_{\rm turb}$ \\
& & & & & & & & & & [mas/yr] & [mas/yr] & [K] & [K] & & [km~s$^{-1}$]\\
\hline\hline
4905854582002922112 & 3 & 0.71819 & $-$60.73536 &  20.3699 & 20.8233 & 19.8899 &  0.9151 &  0.4443  &  0.4708   &  1.712 & $-$3.205 &  5286 & 185 & 3.07 & 1.39 \\ 
4905842693533910912 & 5 & 0.79783  & $-$60.74766 &  17.4662 & 18.0192 & 16.7768 &  1.2245 &  0.5430  &  0.6815   &  0.519 & $-$1.755 &  4658 &  82 & 1.63 & 1.67\\
4905854066606832640 & 6 & 0.60930 & $-$60.77164 &  19.4535 & 19.4474 & 19.4719 &  -0.0445 & -0.0125 & -0.0321   &  0.259 & $-$1.873 & >\,6500 & 259 & 3.53 & 1.29\\
4905842212497084288 & 7 & 0.71638 & $-$60.80579 &  19.1620 & 19.5857 & 18.6837 &  0.8845 &  0.4151  &  0.4694   &  0.546 & $-$1.515 &  5351 & 163 & 2.61 & 1.48\\
4905838158047933056 & 8 & 0.73765 & $-$60.86140 &  19.8914 & 20.2882 & 19.3216 &  0.9498 &  0.3884  &  0.5614   &  1.180 & $-$1.158 &  5190 &  80 & 2.84 & 1.43\\
4905847984933119104 & 9 & 0.43461 & $-$60.83191 &  19.4876 & 19.8772 & 18.9187 &  0.9433 &  0.3820  &  0.5613   &  2.013 & $-$2.812 &  5205 &  79 & 2.69 & 1.46\\
4905841147345181568 & 10 & 0.76487 & $-$60.84038 & 18.8278 & 18.9470 & 18.6767 &  0.2507 &  0.1119  &  0.1388   &  0.701 & $-$1.896 & >\,6500 & 202 & 3.06 & 1.39\\ 
4905842044993754112 & 11 & 0.61753 & $-$60.80400 & 18.9592 & 19.0052 & 18.8705 &  0.1156 &  0.0393  &  0.0763   &  0.604 & $-$1.689 & >\,6500 & 160 & 3.25 & 1.35\\ 
4905839876035355776 & 12 & 0.93365 & $-$60.80132 & 18.9200 & 19.0192 & 18.8233 &  0.1737 &  0.0912  &  0.0825   &  0.577 & $-$1.887 & >\,6500 & 259 & 3.17 & 1.37\\ 
4905838123688194048 & 13 & 0.75174 & $-$60.86246 & 20.2184 & 20.5075 & 19.8392 &  0.6504 &  0.2811  &  0.3693   &  0.090 & $-$3.516 &  5957 & 141 & 3.24 & 1.35\\  
4905647805097805952 & 15 & 0.81263 & $-$60.99411 & 19.0983 & 19.8099 & 18.3984 &  1.3958 &  0.7024  &  0.6934   &  1.568 & $-$2.925 &  4421 & 132 & 2.15 & 1.57\\  
4905842624814433280 & 16 & 0.81460 & $-$60.74894 & 19.8289 & 20.1642 & 19.4524 &  0.6922 &  0.3264  &  0.3658   &  0.624 & $-$2.582 &  5840 & 196 & 3.05 & 1.39\\  
4905836233902550656 & 17 & 0.58846 & $-$60.97393 & 19.4989 & 19.7949 & 19.0049 &  0.7739 &  0.2884  &  0.4854   &  0.459 & $-$1.201 &  5618 &  81 & 2.84 & 1.43\\  
4905843346368941312 & 22 & 1.03213 & $-$60.74080 & 19.7344 & 20.0114 & 19.4214 &  0.5695 &  0.2681  &  0.3014   &  1.225 & $-$2.027 &  6205 & 210 & 3.12 & 1.38\\  
4905839291919803264 & 23 & 1.05756 & $-$60.80160 & 19.8241 & 20.1851 & 19.4789 &  0.6857 &  0.3518  &  0.3339   &  2.820 & $-$3.292 &  5867 & 265 & 3.05 & 1.39\\ 
4905849191820032512 & 113 & 0.49597 &  $-$60.74761 & 17.8764 & 18.2989 & 17.2878 &  0.9946 & 0.4141 &  0.5805   &  1.010 & $-$2.166 &  5094 &  81 & 1.99 & 1.60\\  
4905841632676933376 & 125 & 0.84454 & $-$60.77233 & 20.4191 & 20.6245 & 20.2609 &  0.3427 &  0.1972 &  0.1455   &  $-$0.285 & $-$1.276 & >\,6500 & 339 & 3.61 & 1.28\\  

\hline
 \end{tabular}
\end{sidewaystable}

\clearpage

\begin{table}
\caption{Comparison between stars of Grus~II in our sample and \cite{simon_birds_2020}. The "*" symbol identifies the HB stars. The italic ID identifies the stars whose radial velocity is derived from the blue spectra.}
\label{tab com Simon grus}
\renewcommand{\arraystretch}{1.3}
\footnotesize
\setlength{\tabcolsep}{4pt} 
\begin{adjustbox}{width=\dimexpr\paperwidth-2.5cm\relax}

\begin{tabular}{c c c c c c c c c c c c c}
\hline
 star ID & RA  &   DEC    & v$_{\rm rad}$ &   $\rm e_{v_{rad}}$ & v$_{\rm rad,Simon}$ & e$_{\rm v_{rad,Simon}}$  & Our member & Member for Simon & [Fe/H] & $\sigma_{\rm [Fe/H]}$  & [Fe/H]$_{\rm Simon}$  & $\sigma_{\rm [Fe/H],Simon}$\\
         &     &          & [km~s$^{-1}$] &    [km~s$^{-1}$]    &  [km~s$^{-1}$]      & [km~s$^{-1}$]      &            &                  &        &                   &                       &                       \\
\hline\hline 
 \textit{1}*   & 331.08387 & $-$46.39487   &     $-$98.8  &   0.5   &   $-$108.7   &  2.4   &   $\checkmark$ & $\checkmark$     &    $-$      & $-$    &    $-$    &  $-$   \\ 
 3            & 330.96673 & $-$46.41297   &    $-$107.5  &   0.4   &   $-$109.0   &  1.3   &   $\checkmark$ & $\checkmark$     &  $-$2.93    & 0.12   &   $-$2.99 &  0.28  \\ 
 4            & 331.01845 & $-$46.42266   &    $-$110.3  &   1.4   &    $-$       &   $-$  &   $\checkmark$ &      $-$         &  $-$2.93    & 0.24   &    $-$    &  $-$   \\
 6*           & 331.14065 & $-$46.44445   &    $-$108.0  &   2.1   &    $-$113.2  &  2.1   &   $\checkmark$ & $\checkmark$     &    $-$      & $-$    &    $-$    &  $-$   \\
 8            & 330.88169 & $-$46.53853   &    $-$181.9  &   0.8   &   $-$180.5   &  2.0   &   $\times$     & $\times$         &    $-$      & $-$    &    $-$    &  $-$   \\
 9            & 331.06859 & $-$46.37375   &    $-$110.3  &   2.6   &   $-$110.2   &  2.1   &   $\checkmark$ & $\checkmark$     &  $-$2.92    & 0.49   &   $-$1.82 &  0.33  \\
 10           & 330.80891 & $-$46.51693   &        28.2  &   0.4   &       28.4   &  1.2   &   $\times$     & $\times$         &    $-$      & $-$    &    $-$    &  $-$   \\ 
 13           & 331.07562 & $-$46.54788   &    $-$106.3  &   0.9   &   $-$107.9   &  2.0   &   $\checkmark$ & $\checkmark$     &  $-$2.69    & 0.18   &    $-$    &  $-$   \\ 
 \textit{14}*  & 330.87292 & $-$46.28101   &    $-$104.0  &   1.2   &     $-$      &  $-$   &   $\checkmark$ &      $-$         &    $-$      & $-$    &    $-$    &  $-$   \\ 
 16*          & 331.21856 & $-$46.35556   &    $-$189.7  &   2.3   &     $-$      &  $-$   &   $\times$     &      $-$         &    $-$      & $-$    &    $-$    &  $-$   \\ 
 22*          & 331.02493 & $-$46.48206   &     $-$97.1  &   1.3   &     $-$      &  $-$   &   $\checkmark$ &      $-$         &    $-$      & $-$    &    $-$    &  $-$   \\ 
 23           & 331.10737 & $-$46.25574   &        74.9  &   0.4   &     $-$      &  $-$   &   $\times$     &      $-$         &    $-$      & $-$    &    $-$    &  $-$   \\ 
 24*          & 330.92204 & $-$46.46660   &    $-$112.1  &   1.5   &   $-$107.2   &  2.0   &   $\checkmark$ & $\checkmark$     &    $-$      & $-$    &    $-$    &  $-$   \\ 
 25           & 331.03710 & $-$46.32437   &        30.1  &   1.1   &       31.1   &  2.7   &   $\times$     & $\times$         &    $-$      & $-$    &    $-$    &  $-$   \\ 
 26           & 331.04160 & $-$46.35068   &    $-$107.0  &   0.3   &   $-$108.0   &  1.1   &   $\checkmark$ & $\checkmark$     &  $-$2.76    & 0.07   &   $-$2.76 &  0.17  \\  
 27*          & 330.99526 & $-$46.57928   &       171.1  &   2.1   &    $-$       &  $-$   &   $\times$     &      $-$         &    $-$      & $-$    &    $-$    &  $-$   \\ 
 29           & 331.00051 & $-$46.42478   &     $-$98.3  &   1.3   &   $-$110.9   &  1.3   &   $\checkmark$ & $\times$         &  $-$2.24    & 0.24   &    $-$    &  $-$   \\
 30           & 330.99051 & $-$46.43441   &    $-$108.7  &   0.7   &   $-$109.6   &  1.3   &   $\checkmark$ & $\checkmark$     &  $-$2.88    & 0.17   &   $-$2.76 &  0.25  \\  
 \textit{114}* & 331.30712 & $-$46.33453   &     $-$22.4  &   1.9   &     $-$      &  $-$   &   $\times$     &     $-$          &    $-$      & $-$    &    $-$    &  $-$   \\ 
 121          & 331.17323 & $-$46.37917   &    $-$158.6  &   0.2   &   $-$160.8   &  1.0   &   $\times$     & $\times$         &    $-$      & $-$    &    $-$    &  $-$   \\ 
 122          & 331.09963 & $-$46.61741   &    $-$105.7  &   0.1   &   $-$108.6   &  1.0   &   $\checkmark$ & $\checkmark$     &  $-$2.47    & 0.05   &   $-$2.62 &  0.16  \\ 
\hline

 \end{tabular}
 \end{adjustbox}
\end{table}

\begin{table}
\caption{Comparison between stars of Tucana~IV in our sample and \cite{simon_birds_2020}. The "*" symbol identifies the HB stars. The $\triangle$ represents the "unlikely member" stars.}
\label{tab com Simon tuc}
\renewcommand{\arraystretch}{1.3}
\centering
\footnotesize
\setlength{\tabcolsep}{4pt} 
\begin{adjustbox}{width=\dimexpr\paperwidth-2.5cm\relax}

\begin{tabular}{c c c c c c c c c c c c c}
\hline
 star ID & RA  &   DEC   & v$_{\rm rad}$ & e$_{v_{rad}}$  & v$_{\rm rad,Simon}$ & e$_{\rm v_{rad,Simon}}$ & Our member & Member for Simon & [Fe/H] & $\sigma_{\rm [Fe/H]}$ & [Fe/H]$_{\rm Simon}$ & $\sigma_{\rm [Fe/H],Simon}$\\
         &     &         &[km~s$^{-1}$]  & [km~s$^{-1}$]  &  [km~s$^{-1}$]      &   [km~s$^{-1}$]         &            &                  &        &                  &                      &                       \\
\hline\hline
 3              &   0.71819      & $-$60.73536   &       164.4    &  2.1    &    162.0   &  0.70  & $\times$     & $\times$      &    $-$    &   $-$   &  $-$      &  $-$\\ 
 5              &   0.79783      & $-$60.74766   &        16.6    &  0.2    &     14.7   &  1.1   & $\checkmark$ &      $-$      &  $-$1.38  &  0.07   &  $-$      &  $-$\\
 6*             &   0.60930      & $-$60.77164   &         5.6    &  2.1    &      $-$   &  $-$   & $\checkmark$ &      $-$      &    $-$    &   $-$   &  $-$      &  $-$\\
 7              &   0.71638      & $-$60.80579   &        13.5    &  0.7    &      9.8   &  1.5   & $\checkmark$ & $\checkmark$  &  $-$2.75  &  0.17   &  $-$2.59  & 0.30\\
 8              &   0.73765      & $-$60.86140   &        16.9    &  1.4    &     16.5   &  2.2   & $\checkmark$ & $\checkmark$  &  $-$2.09  &  0.25   &  $-$2.20  & 0.32\\
 9              &   0.43461      & $-$60.83191   &        29.3    &  1.3    &     31.7   &  7.3   & $\triangle$  & $\times$      &  $-$1.28  &  0.24   &  $-$      &  $-$\\  
 10*            &   0.76487      & $-$60.84038   &        19.6    &  1.0    &     18.3   &  2.5   & $\checkmark$ & $\checkmark$  &    $-$    &  $-$    &  $-$      &  $-$\\ 
 11*            &   0.61753      & $-$60.80400   &        14.0    &  1.3    &     13.5   &  3.3   & $\checkmark$ & $\checkmark$  &    $-$    &  $-$    &  $-$      &  $-$\\ 
 12*            &   0.93365      & $-$60.80132   &        20.5    &  1.5    &     21.1   &  2.8   & $\checkmark$ & $\checkmark$  &    $-$    &  $-$    &  $-$      &  $-$\\ 
 13             &   0.75174      & $-$60.86246   &    $-$203.8    &  1.8    & $-$205.4   &  5.2   & $\times$     & $\times$      &    $-$    &  $-$    &  $-$      &  $-$\\  
 15             &   0.81263      & $-$60.99411   &       143.3    &  1.1    &      $-$   &  $-$   & $\times$     &      $-$      &    $-$    &  $-$    &  $-$      &  $-$\\  
 16             &   0.81460      & $-$60.74894   &     $-$18.1    &  1.7    &  $-$19.3   &  2.6   & $\triangle$      & $\times$      &  $-$2.46  &  0.29   &  $-$      &  $-$\\ 
 17             &   0.58846      & $-$60.97393   &        59.3    &  1.7    &      $-$   &  $-$   & $\triangle$  &      $-$      &  $-$2.32  &  0.29   &  $-$      &  $-$\\ 
 22             &   1.03213      & $-$60.74080   &        60.3    &  1.7    &      $-$   &  $-$   & $\triangle$      &      $-$      &  $-$1.86  &  0.30   &  $-$      &  $-$\\  
 23             &   1.05756      & $-$60.80160   &       125.3    &  1.9    &      $-$   &  $-$   & $\times$     &      $-$      &    $-$    &  $-$    &  $-$      &  $-$\\  
 113            &   0.49597      & $-$60.74761   &        43.5    &  0.4    &      $-$   &  $-$   & $\triangle$  &      $-$      &  $-$1.48  &  0.11   &  $-$      &  $-$\\ 
 \textit{125}*  &   0.84454      & $-$60.77233   &     $-$40.0    &  2.5    &      $-$   &  $-$   & $\triangle$  &      $-$      &    $-$    &  $-$    &  $-$      &  $-$\\
 
\hline

 \end{tabular}
  \end{adjustbox}

\end{table}

\end{document}